\documentclass[11pt]{article}
\usepackage[utf8]{inputenc}
\usepackage[T1]{fontenc}
\usepackage{lmodern}
\usepackage{textcomp}
\usepackage{microtype}

\usepackage{newunicodechar}

\usepackage[margin=0.9in, top=0.85in]{geometry}

\renewenvironment{abstract}{%
  \small
  \begin{center}\bfseries\abstractname\end{center}%
  \setlength{\parskip}{0pt}%
  \setlength{\parindent}{1.5em}%
  \leftskip=0.75em \rightskip=0.75em
  \par
}{\par\medskip}
\usepackage{setspace}
\usepackage{parskip}

\usepackage{amsmath, amssymb}
\usepackage{graphicx}
\graphicspath{{figures/}{../paper/figures/}}
\usepackage{booktabs}
\usepackage{array}
\usepackage{longtable}
\usepackage{enumitem}
\usepackage{xcolor}
\usepackage[colorlinks=true, linkcolor=blue, citecolor=blue, urlcolor=blue]{hyperref}
\usepackage[round, sort&compress]{natbib}
\usepackage{lineno}
\usepackage{titlesec}
\usepackage{caption}
\usepackage{subcaption}
\usepackage[most]{tcolorbox}
\usepackage{placeins}
\usepackage{float}

\newcounter{conceptbox}
\renewcommand{\theconceptbox}{\arabic{conceptbox}}
\newenvironment{conceptbox}[1][]
  {\par\medskip
   \begin{tcolorbox}[
     enhanced, colback=white, colframe=black!55,
     boxrule=0.5pt, arc=2pt, left=8pt, right=8pt, top=6pt, bottom=6pt,
     title={\textbf{Box \theconceptbox.} #1},
     coltitle=black, colbacktitle=black!4, fonttitle=\normalfont,
     attach boxed title to top left={xshift=8pt, yshift*=-2pt},
     boxed title style={colback=black!4, colframe=black!55, sharp corners, boxrule=0.4pt}
   ]}
  {\end{tcolorbox}\par\medskip}

\newcolumntype{L}[1]{>{\raggedright\arraybackslash}p{#1}}
\newcolumntype{C}[1]{>{\centering\arraybackslash}p{#1}}
\newcolumntype{R}[1]{>{\raggedleft\arraybackslash}p{#1}}

\titleformat{\section}{\Large\bfseries}{\thesection}{0.9em}{}
\titleformat{\subsection}{\large\bfseries}{\thesubsection}{0.8em}{}
\titleformat{\subsubsection}{\normalsize\bfseries\itshape}{\thesubsubsection}{0.75em}{}
\titlespacing{\section}{0pt}{1.4em}{0.5em}
\titlespacing{\subsection}{0pt}{1.0em}{0.35em}
\titlespacing{\subsubsection}{0pt}{0.85em}{0.25em}

\newcommand{\eci}{\textsc{eci}}
\newcommand{\fl}{\textsc{frontierlag}}
\newcommand{\versio}{\textsc{versio-ai}}

\title{%
Frontier Lag:\\
A Bibliometric Audit of Capability Misrepresentation\thanks{In the title, we use the term \textit{misrepresentation} to refer to a broad pattern of mismatched claims and scope across our corpus. We do not mean to imply that individual researchers are acting in bad faith, or that the misrepresentation is intentional. Our criticisms are directed at reporting norms and incentive structures in the field.}\\
in Academic AI Evaluation%
}

\author{%
  David Gringras\\[2pt]
  \small Harvard University\\[-2pt]
  \small \texttt{davidgringras@hsph.harvard.edu}
  \and
  Misha Salahshoor\\[2pt]
  \small AISST, Cambridge, MA\\[-2pt]
  \small \texttt{misha@cbai.ai}
}

\date{September 2026}\hypersetup{pdftitle={Frontier Lag: A Bibliometric Audit of Capability Misrepresentation in Academic AI Evaluation}, pdfauthor={David Gringras, Misha Salahshoor}}

\begin{document}
\maketitle
\vspace{-1.5em}

\begin{abstract}

\setlength{\baselineskip}{0.97\baselineskip}

Evaluations of large language models (LLMs) in applied domains often test models that have already been surpassed by the time the evaluation is published. In this paper, we define the publication elicitation gap as a measure of the difference between the set of AI systems whose performance is reported in a published paper, and the set of AI systems which a reasonable reader today would understand the paper to be talking about. We conduct a systematic search of the OpenAlex database for papers published between 2022-01-01 and 2026-04-01 with LLM-related keywords (n = 112,303) and identify papers that evaluate LLMs (n = 18,574 admissible records). We then rank the LLMs evaluated in these papers using the Epoch AI Capabilities Index (\eci{}), an aggregate measure of LLM capability, relative to a ``frontier LLM'', i.e. the highest-\eci{} LLM publicly available at the time when the LLMs were evaluated. We find that the median paper's models are worse than the frontier LLM at the time of evaluation (a median gap of +10.45 \eci{}; H1, n = 12,668). The gap is increasing at a rate of +4.07 \eci{} per year (H2, nominal 95\% CI [+3.75, +4.45]). The sign of the H1 result is unchanged if we do not impute any evaluation dates. For the n = 746 papers which explicitly state their evaluation date and in which the model tested has an \eci{} score, the median gap for H1 is +3.52 \eci{}. Only 18.4\% of full-text papers specify the date of evaluation explicitly. A Bayes-corrected 52.5\% (95\% CI: [47.3, 57.9]) of the abstracts audited discuss their conclusions in terms of ``AI'' as a category, rather than specific models. Just 2.2\% of abstracts and 21.2\% of full-text articles on reasoning models disclose whether the models were evaluated with reasoning turned on or off (H4). To address these gaps, we propose (a) a reporting standard for authors (we propose \versio{} v1.2, a 13-item checklist based on our configuration surface), (b) enforcement of this standard by journal editors and peer reviewers, and (c) the conditioning of grant funding on such disclosure and the provision of API access by funders.

\end{abstract}

\section{Introduction}
\label{sec:introduction}

\subsection{The structural observation}
\label{sec:intro:problem}

From medicine to law, coding to education and scientific reasoning, a large array of research papers focus on evaluating LLMs. If we read the research literature globally, we can be misled about the current capabilities of AI systems in these respective areas. While these papers measure the performance of specific models, they often do so in a way that does not correspond to the claims made in their titles and abstracts about ``AI'' as a general category. In many cases, the models being evaluated are older or lower-tier, as when a paper written in 2026 tests GPT-4o-mini zero-shot, while the frontier of AI has moved on to new reasoning-capable, tool-using models such as GPT-5.5 Pro and Claude Opus 4.7.

The methods sections of the papers we audit usually do not fully document model tier and elicitation, and the claims often made in the abstracts about ``AI'' can be cited by clinical, legal, and policy papers. We measure the frequency of mismatch between abstract claims and the models and elicitation documented in these papers. The pre-registered candidate pool contains 112,303 OpenAlex results matching LLM keywords and dated 2022-01-01 to 2026-04-01; 18,574 papers pass the inclusion gate. The \eci{} gap, our measure of distance from the frontier, is only the observable trace of this mismatch. The median paper in our data ($n = 12{,}668$) evaluates a model which is about 1.3 times as distant from the contemporaneous frontier as Claude 3.7 Sonnet is from Claude Opus 4.5 (142.0 and 149.9 \eci{}; $10.45 / 7.9 = 1.32$), models from the same company that are one major version and one model tier apart. This gap is growing.

Frontier AI labs release new models every few weeks or months, whereas new academic papers come out every year or so. In the time between when a paper conducts its experiments and when its abstract is cited as prior work, the major AI labs will likely have released newer models, along with new reasoning toggles, tool harnesses, and agentic scaffolding. And so the results reported in the paper will reflect the capabilities of an AI system that is outdated compared to the contemporaneous AI frontier, along the dimensions not tested for in the paper.

This gap is broken down into three components, which are tracked individually in the audit. The first is temporal lag, which is defined as the time between the release date of the model tested and the frontier at the time of testing. The second is tier lag: a sibling model from the same model family with a higher \eci{} was released by the time of testing. This applies only if the tested model and the higher-tier model were released within 90 days of each other. For example, GPT-4o-mini was tested while GPT-4o was available, or Claude 3 Sonnet was tested while Claude 3 Opus was available, or Gemini 1.5 Flash while Gemini 1.5 Pro was available. The third is configuration underspecification: that is, what settings the original paper did or did not report, like the use of reasoning modes, tools, scaffolding, the sampling temperature, and prompt design. In the papers we audit, researchers typically report one `headline' performance figure, which combines these three elements, and readers can compound this by interpreting claims about one model as claims about `AI'. We do not propose any `overall' combination of these different elements of the gap, and leave it to the reader to decide how to weight them.

Of these three elements, `configuration underspecification' has the clearest precedents in the related literature. Researchers at Apollo Research, for instance, refer to the model-level version of this configuration underspecification as the `evals gap' \citep{apolloresearch2024evalsgap}. This is the gap between what models can demonstrate, and what they demonstrate to a naive tester. \citet{hochlehnert2025sober} quantify this effect in the context of mathematical reasoning, showing how the rankings of models' `state of the art' performance can change depending on the decoding parameters, seeds, prompt formatting, and hardware that is used. Closer to this study is \citet{balloccu2024leak}, who audited 255 studies that had used the ChatGPT interface for testing, drawing on a pre-existing taxonomy of data contamination and malpractice `signals'. They find that around 4.7m benchmark test samples were tested against models within a single year. We build on this approach by introducing an explicit capability-distance based framework for AI audit studies, which we implement across five fields and four years and ground in a pre-registered analysis plan.

Suppose that a paper published in 2026 tests free-tier ChatGPT, using zero-shot prompting, no tool use, no reasoning modes, and no scaffolded comparison to other models. This paper is behind the contemporaneous frontier on all these dimensions introduced by the new `reasoning-era' systems, and these effects can compound. If the paper tests a tier-two model using no reasoning and no tool use, and was evaluated in 2024 but cited in 2026, this means it is describing the capabilities of a 2023 product to users who have access to a 2026 product. Even if extra inference-time compute can make up for some of the difference between having and not having multi-agent scaffolding, each element of this gap limits what can be recovered by other elements. We call the interaction between these factors the publication elicitation gap, which is potentially affected by the timing of publication, API-access considerations, and reporting practices. This is a version, in the academic literature, of the `elicitation gap' that the Apollo/METR/AISI programme of research has documented at the model level.

\begin{figure}[!t]
\refstepcounter{conceptbox}\label{fig:box1}%
\begin{conceptbox}[Two lenses on compound capability failure]
\centering
\includegraphics[width=\linewidth]{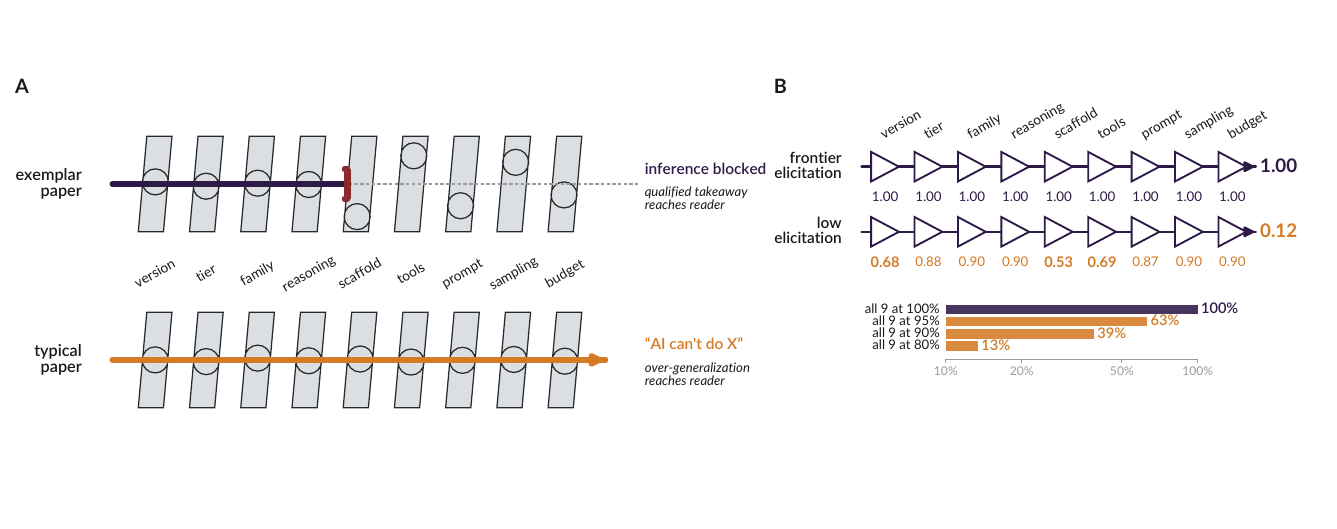}

\smallskip
\footnotesize Panel A: Reporting and elicitation safeguards against overgeneralisation from single-configuration results, represented using Reason's Swiss-cheese model of compound causation \citep{reason2000}. Each slice of cheese represents a safeguard, and each hole represents a failure. The arrow depicts the inference that readers might make from an under-specified abstract to a class-level claim (e.g., ``AI can't do X''). Top row: Exemplar cheese model, with holes scattered such that at least one safeguard prevents an overgeneralised inference (in this example, scaffolding). Bottom row: Schematic, with all failures coinciding such that the overgeneralised inference makes it through. One explanation for such a pattern could be correlated cost, access, and reporting constraints, although we did not test for this mechanism. Panel B: A mathematical restatement of the same compound-causation schematic as in A, with example numbers anchored to a published score on SWE-Bench-Verified (80.8\% pass@1, Anthropic Opus 4.6 Thinking Max, mean over 25 trials). The numbers we present as fractions retained are derived from public comparisons and interpolated bounds (see Figure~\ref{fig:waterfall} and Table~S2). If taken as independent, these fractions yield an example compound capability $G_\mathrm{total} \approx 0.12$ of frontier-equivalent capability. The bars in Panel B show what happens when factors are uniformly attenuated: for example, nine axes at 95\% leaves 63\% total capability, and nine axes at 90\% leaves 39\%. Note that independence is a simplification: in real-world usage, elicitation axes interact and can compensate for one another. Note also that four factors are confounded with model changes beyond the axis suggested by their label. In the reasoning (0.90) factor, we also change model from Opus 4.6 to Opus 4. In the tier (0.88) factor, we also change model from Opus 4 to the previous-generation 3.7 Sonnet. In the scaffold (0.53) factor, we also change model from Claude 3.7 Sonnet to Claude 3.5 Sonnet. And in the tools (0.69) factor, we also change model from Claude 3.5 Sonnet to GPT-4o, holding the scaffold fixed to SWE-agent.
\end{conceptbox}
\end{figure}

\subsection{The structural reframe}
\label{sec:intro:reframe}

The most common way that the abstracts of AI evaluation papers describe the scope of their results is to make a class-level generalisation from a finding about a particular model to the entire class of ``AI.'' We pre-registered a Bayes-corrected estimation technique, and found that the rate of class-level claims is 52.5\% (95\% confidence interval 47.3\% to 57.9\%). We also find that, if we do not correct, the odds of an abstract containing a class-level claim increase significantly with year of publication over the whole V4F-cascaded corpus (odds ratio 1.23; 95\% confidence interval 1.19 to 1.27; p<$10^{-33}$; $n = 18{,}565$ abstracts). This significant increase is also present when we restrict our sample to those papers which are covered by the Chatbot Arena Elo scores, or by the Artificial Analysis intelligence index. These increasing odds show that the most common interpretation of scope is only becoming more entrenched over time.

Consider the following hypothetical composite from the 2026 cohort: a clinical study testing GPT-4o mini under spare elicitation (zero-shot prompting, no reasoning option, no tool use, no web-search) without reporting the model version or test-harness used in the methods section and making class-level claims in the abstract (``LLMs fail at clinical diagnosis''). The weekend after the paper was published, a news article in the science section could read ``AI fails at clinical diagnosis'' and propagate into the clinical/regulatory/policy citation chain. A clinician, after reading the abstract and headline, may update towards believing that LLMs are unable to do clinical diagnosis. Given the information in the abstract, that would be a reasonable conclusion, and it is also reasonable that the authors reported the findings of their study. But something is lost in translation: ``GPT-4o mini, reasoning off, no tools, no web search.'' That missing qualifier is the focus of the present paper.

While we describe the models tested in the literature and the scope of the claims made, we do not know why exactly researchers chose to test the models they did. Similarly, we do not know if any particular result would be different if it were tested again using contemporaneous frontier models. There are a variety of factors that may explain the pattern of results we observe, such as publication delays or lack of model access. Future work is needed to weigh these (and other) possibilities; such efforts will require more information than is provided by the gap distribution and external lag benchmark alone. Regardless of the explanation, our pre-registered decision to avoid identifying any negative exemplars (i.e., pointing to any specific paper as a ``bad example'') is unaffected.

\subsection{What this paper does}
\label{sec:intro:contribution}

We start with a pre-registered set of 112,303 candidate records from OpenAlex matching LLM keywords for the date range 2022-01-01 to 2026-04-01. Of these, we retain only those that pass the gate of our pre-registered inclusion classifier ($n = 18{,}574$ records) for our primary analyses (Section~\ref{sec:methods:corpus}). To measure how far each model evaluated is from the frontier, we rely primarily on one scale: the Epoch AI Capabilities Index (\eci{}; \citealp{epochai2024}), which is anchored to GPT-5 in August 2025 with a value of 150 and calibrated using data from $\sim$165 models on both sides of the frontier/near-frontier line. We also rely on two sensitivity scales: Chatbot Arena Elo and the Artificial Analysis intelligence index.

We reject the null for our pre-registered confirmatory hypotheses H1 (location) and H3 (tier lag), but not for H6 (valence asymmetry). Because only papers with a sibling with a higher \eci{} are considered for H3, we interpret the magnitude of H3 conditional on that selection effect. Rejection of the null is not sufficient to show that H3 holds. Our primary descriptive headline result is the compound failure rate, H5. The full set of tests, pre-registered framing maps, and descriptive magnitudes is in Section~\ref{sec:methods}.\footnote{Per-paper component inputs are in the data bundle, available from the authors on request; corpus-level component magnitudes coincide with the H1, H3, and H4 primaries reported in \S\ref{sec:results:confirmatory}, so a separate corpus-level decomposition table would duplicate them.}

Appendix~\ref{app:versio-ai} additionally contains a reporting checklist, \versio{} v1.2 (\url{https://doi.org/10.5281/zenodo.20060459}). This reporting checklist is a candidate specification, and it is open to community comment for 60 days. It is intended to be used in conjunction with existing reporting standards such as CONSORT-AI and SPIRIT-AI for clinical trials and clinical trial protocols \citep{liu2020consort, cruzrivera2020spirit}; TRIPOD+AI and TRIPOD-LLM for clinical prediction models \citep{collins2024tripodai, gallifant2025tripod}, where TRIPOD-LLM is the LLM-specific version with 19 main items and 50 subitems; DECIDE-AI for early-stage evaluations of clinical decision support systems \citep{vasey2022decide}; and STARD-AI for diagnostic accuracy studies \citep{sounderajah2025stardai}. None of these frameworks apply to the most common type of capability evaluation: an off-the-shelf empirical probe of a named LLM on an applied task, with no prediction model involved and no randomised controlled trial. \versio{} is intended to be integrated into these reporting frameworks, such that its specification of the elicitation-surface items is a small add-on to the relevant framework for the type of evaluation, and the standalone document is a catch-all for capability evaluations not covered by other frameworks.

The third and final artefact is \fl{} (\url{https://frontierlag.org}), an accompanying Python package and live web tool (built on the same backend as the audit; \url{https://pypi.org/project/frontierlag/}, \url{https://doi.org/10.5281/zenodo.20060457}). Users can paste a DOI to retrieve its audit report if it is covered by the frozen dataset used here. DOIs outside the corpus are resolved live via CrossRef and OpenAlex. We pre-registered the study on the Open Science Framework (\url{https://doi.org/10.17605/OSF.IO/7XM3D}) before depositing on arXiv.

\subsection{What this paper does not do}
\label{sec:intro:scope}

Distance from the frontier is not distance from truth. While this work measures structural lag at the level of the literature, future work is needed to see if the composite case in \S\ref{sec:intro:reframe} would still report failure with a frontier model using reasoning in a tool-use harness. We also leave re-running evaluations in such a manner, a form of replication, to future work. \versio{} has similar limitations: the checklist evaluates what authors explicitly reported, but cannot directly assess what they did not. We do not know if undisclosed information, had it been elicited and reported, would have changed the outcome. Similarly, we do not know if any given abstract is ``wrong'' about a given result: our audit process describes what the abstract commits to, and replication determines whether those commitments hold.

We do not name specific papers as negative exemplars; instead, our critique is aimed at the environment those papers operated in. NEJM AI's first issue was published in January 2024, and manuscripts in it were submitted in early 2023, when the frontier model was GPT-3.5 and the recently released GPT-4: there was no ``reasoning'' dial to turn off, there was no o1. But by May 2026, all those authors, reviewers, and publication cycles operate downstream of GPT-5.5 Pro, Claude Opus 4.7, and Gemini 3.1 Pro: models that support reasoning toggles and harnessed tool-use and budget-tier API access, none previously available. We tabulate six articles, read as in-scope under \versio{} v1.2 criteria, as positive exemplars in Appendix~H. Our choice to tabulate positive examples by-name, but not the aggregate corpus, was pre-registered.

Throughout the rest of the paper, we present two sets of three categories each. Our measurement pipeline scores in-scope articles by capability, elicitation, and interpretation, while our analysis conceptualises the gap in terms of temporal lag, tier lag, and configuration underspecification. These two sets of categories do not map symmetrically onto one another: ``capability'' collapses temporal and tier lag into one measured category, while ``configuration'' is just elicitation by another name, and the ``interpretive'' category is about the reader. Any failures in the first two should have been attenuated by the time of abstract authorship. H1, H3, and H6 depend on this asymmetry, which would be obscured by a single six-cell taxonomy.

\section{Background and related work}
\label{sec:background}

\subsection{Evals gap and elicitation gap}
\label{sec:background:gaps}

A claim about a model's capabilities without an accompanying claim about the elicitation surface used is really just a claim about a testing configuration. And this is the typical case in the capabilities literature. This issue of underspecification in model capability claims has been diagnosed by Apollo Research, METR, and the UK AISI. Apollo Research's ``evals gap'' \citep{apolloresearch2024evalsgap} is the observation that models which fail in zero-shot settings without any tools or scaffolding often succeed when tested with better elicitation surfaces. Meanwhile, METR has demonstrated this empirically for certain agentic tasks, showing that differences in capabilities resulting from different elicitation are much larger than differences across model versions \citep{metr2024elicitation}. And AISI's ``frontier-trends'' evaluations treat AI capabilities evaluations as an evolving challenge, where what matters is understanding the trajectory across models, tiers, and elicitation surfaces, and where the results of any particular benchmark are only meaningful as a snapshot in time \citep{aisi2025frontiertrends}. The mechanism described by Apollo, METR, and the AISI is a model-level elicitation gap. We refer to the reflection of this model-level elicitation gap in the academic literature as the publication elicitation gap, which may be affected by publication timing, API access, and reporting practices. In addition to estimating the prevalence of the publication elicitation gap across five applied domains, we audit the availability of information in the academic literature necessary to situate capability claims along this trajectory.

\subsection{Measurement templates from AI safety research}
\label{sec:background:templates}

The most similar methodology to this audit is the Safetywashing audit by CAIS \citep{cais2024safetywashing}, which audits the difference between claims made about AI safety and the evidence behind those claims, across a corpus aggregated by benchmarks, defines a construct, releases code, and proposes solutions through reporting discipline. Just as CAIS targets safety-claim inflation, we target capability-claim mislocation. The mechanisms behind both problems are similar, hinging on the difference between claims and the evidence that underlies them, and the importance of reporting discipline to prevent misleading claims spreading down the line. The corpora and solutions differ: CAIS focuses on safety benchmarks and claims about safety, while we focus on applied-domain capability evaluations and reporting standards for model versions and elicitation methods. \versio{} is the reporting-standards sibling to their safetywashing construct.

\citet{bean2025measuring} is the most similar in scale to this paper as a contemporary example of a methodological audit. The authors reviewed 445 benchmarks for LLMs published in top conferences, recommending 8 design improvements for benchmark validity. The subject of their audit is upstream to the subject of this audit. They ask whether HumanEval is a valid measure of coding ability, and so on for the 444 other benchmarks they study. We ask the analogous question about the published evaluations using HumanEval: whether the reported result for GPT-3.5 in a given publication informs the reader about what AI can currently do. \versio{} builds on Bean et al.'s recommendations for benchmark design, taking them a step further to reporting standards for evaluations published in the literature.

The 2026 AI Index \citep{hai2026aiindex} documents the same trend, longitudinally, at the level of the field: the average score on the Foundation Model Transparency Index has declined from 58 to 40 over the past two releases. The complexity of the configuration space to be captured by the literature is increasing faster than reporting practices can keep up. Our audit documents the same phenomenon at the level of individual papers.

\subsection{Prior bibliometric audits}
\label{sec:background:priors}

\citet{nagendran2020clinician} present a bibliometric audit ancestor in the clinical AI space, comparing the performance of deep learning systems and clinicians, establishing a template for rigorous capability claim audits with systematic sampling of an identified corpus of research, pre-registered quality criteria, and domain-stratified (rather than pooled) outcomes. Our corpus follows this template, though expands scope to law, coding, education, and scientific reasoning, and adds an additional measurement layer on top of previous bibliometric audits: capability distance.

Most proximally, \citet{balloccu2024leak} conduct an audit of 255 studies evaluating ChatGPT, finding roughly 4.7 million benchmark samples exposed over the course of a year to the models under consideration, using a systematic corpus with explicit coding of methods along a taxonomy of contamination and other evaluation malpractice signals. We follow and extend their structural template, but with new analyses within a capability distance framework and with pre-registered decision rules. The ``evaluation illusion'' described by \citet{agrawal2025evaluation} in the medical LLM literature is conceptually similar to our H6 valence-asymmetry hypothesis, though they do not assess the three-factor decomposition. The most direct precursor to our approach of LLM-powered extraction and coding of articles is \citet{briggs2025llmtrack}, who evaluate 2,674 political science articles with a frontier LLM extraction pipeline, reconciling LLM outputs with human-coded data, with disagreements adjudicated by the team's leadership. Our two-stage V4F pipeline is indebted to Briggs et al.'s approach. \citet{kapoor2023leakage} discuss the reproducibility crisis in sciences that use ML. We extend this framing to the provenance of capability claims, which is the crisis when evaluators elide the version-and-configuration surface.

The most recent prior study similar to our work is the bibliometric review by \citet{chen2026natmed}, which examines LLM evaluation practice in clinical medicine. Chen et al. used GPT-5 reasoning-high calibrated against human reviewers (sensitivity 0.911; specificity 0.921; $\kappa$=0.820) to identify, from 12,894 deduplicated PubMed, Embase, and Scopus records, 4,609 LLM evaluation articles (screened from Jan 2022 through Sept 2025) to include in their analysis. Chen et al. identified nineteen prospective randomised controlled trials (RCTs) in their corpus. Their analysis also showed that 65.7\% of evaluated models were ChatGPT or OpenAI models. Both our audit and Chen et al. use an LLM-assisted screening approach calibrated against a human-coded validation set. However, the focus of our analyses is different. Where Chen et al. study the realism of experimental setups used in the LLMs-in-medicine literature, finding a dearth of prospective RCTs, our audit instead focuses on the gap between models evaluated in the LLM evaluation literature and the contemporaneous frontier, the elicitation surface recorded (or not) in the evaluation, and the framing of the results in paper abstracts. Chen et al. is the closest single-domain comparator to our work in terms of corpus size: Chen et al. identified 4,609 papers on LLMs in medicine, compared to our 6,108 medicine-domain papers (about 2,400 with retrievable full text).

\citet{ko2025miclear} use the MI-CLEAR-LLM checklist to audit 159 medical LLM papers in top-decile journals (Nov 2022 through Jun 2024). Like our audit, they find that easy-to-satisfy checklist items are reported more often (model version: 96.9\%; training data cutoff: 54.1\%), but important details that would enable readers to make sense of results are not (web access: 6.3\%; query date: 50.9\%; exact prompt: 49.1\%; stochasticity: 15.1\%). Ko et al. is the closest single-domain precedent for our H4 and H5 disclosure measurements. Our work extends the analysis of Ko et al. to a corpus roughly two orders of magnitude larger, as well as to five domains. We also provide additional layers of measurement (capability distance to a contemporaneous frontier, within-family model tier lag, and class-level conclusion framing) that are not measured by MI-CLEAR-LLM. The only measurement in common between our works (on overlapping populations) is query date/evaluation date, reported as present in 50.9\% of cases in Ko et al., compared to 18.4\% in the full-text subset of our corpus. The Ko et al. population is limited to exclusively top-decile journals, whereas our population consists of mixed journal tiers by design. Given this difference, journal tier selection would be expected to predict the direction observed in Ko et al.

A later preprint by \citet{bintareaf2026widening}, posted on 10 September 2026 after the first public version of this audit (5 May 2026), maps $11{,}628$ PubMed records on large language models in medicine from January 2023 to June 2026 and reports a widening evaluation lag, measured as the time from ``the release of the newest model a study names'' to the study's publication. Whereas Bin Tareaf et al. consider release age on a single-domain corpus (PubMed), our audit considers capability-relative distance on a cross-domain corpus (OpenAlex) at the evaluation date. Both studies identify a widening lag using these different measures on different corpora; neither replicates the other.

\subsection{Reporting guidelines}
\label{sec:background:guidelines}

The reporting-guideline ecology around AI in applied domains is richer than the conversation around AI evaluation often lets on. That said, none of these frameworks covers the scope that \versio{} targets. For example, CONSORT-AI \citep{liu2020consort} and SPIRIT-AI \citep{cruzrivera2020spirit} are reporting guidelines for clinical trials of AI interventions and their associated trial protocols. That is interventional clinical research, not off-the-shelf capability evaluation. TRIPOD+AI \citep{collins2024tripodai} and TRIPOD-LLM \citep{gallifant2025tripod} are reporting guidelines for clinical prediction model development and reporting. TRIPOD-LLM is explicitly LLM-oriented, with 19 main items and 50 subitems. This is a related but distinct topic that only partially overlaps with the modal applied-domain capability evaluation. DECIDE-AI \citep{vasey2022decide} is a reporting guideline for early-stage evaluations of decision support systems based on AI; this is a tighter scope than the modal off-the-shelf evaluation paper. STARD-AI \citep{sounderajah2025stardai} is a reporting guideline for studies evaluating the diagnostic accuracy of AI. That is out of scope for the modal capability eval.

These reporting frameworks describe neighbouring forms of evaluation (clinical trials, prediction model development, decision support evaluation, and diagnostic accuracy testing) that do not capture the off-the-shelf capability evaluation targeted by \versio{}. Of these frameworks, TRIPOD-LLM has the most scope-overlap with \versio{}: with its 19 items, it expands the reporting of prediction model development (including model development pipeline, model validation, and intended use) to the LLM context. The elicitation surface, which determines what specific LLM is being assessed at a given point in time, is still out-of-scope for TRIPOD-LLM, as it is for all these frameworks. The elicitation surface includes factors such as the model snapshot, the date of evaluation, reasoning mode, tool access, scaffolding, prompting strategy, and sampling parameters. Any of these reporting frameworks could incorporate the 7 fields \versio{} adds as a small extension, at low cost to their committees. The ideal outcome for the \versio{} checklist is for it to be integrated into the relevant reporting framework for the type of evaluation being conducted. \versio{} would then only be a standalone document for evaluations that are not covered by a reporting framework. In cases where the \versio{} scope does overlap with reporting frameworks (for example, a TRIPOD-LLM-scoped paper that also evaluates an LLM at the elicitation surface), both should be cited.

Beyond the tradition of medical and clinical-prediction reporting guidelines, the software engineering research community has been converging on a similar framework. \citet{baltes2026guidelines} develop a reporting checklist for using LLMs in software engineering research. The work has 22 co-authors and covers 8 reporting items. These items include LLM-usage declaration, model versions and configurations, LLM tool architecture, input prompts and logs, human validation, baseline models, metrics used, and limitations. The \citet{baltes2026guidelines} reporting checklist addresses the same challenge as \versio{}: the non-deterministic nature of model output, the opaque composition of LLM training data, and the rapid pace of LLM version changes (usually on the order of weeks) limit the reproducibility of LLM-based measurements. The checklist addresses overlapping concerns with \versio{}, but at a different scope. \citet{baltes2026guidelines} governs software engineering research that uses LLMs as a tool, while \versio{} governs published capability evaluation claims regardless of discipline. The overlap between the two comes across as convergent evolution on a similar problem structure, rather than a coordination outcome.

At a higher level of generality, \citet{kapoor2024reforms} introduce REFORMS, a 32-item reporting checklist for ML-based science. This work draws on 19 researchers and is published in Science Advances. \versio{} can be seen as a domain specialisation of the REFORMS reporting framework, where the general ML reporting items in REFORMS take the specific form of the elicitation surface factors motivated by the audit, rather than running on a parallel track. Those factors include model snapshot, date of evaluation, access tier, reasoning mode and effort, and tool access and scaffolding.

\subsection{Human comparator design literature}
\label{sec:background:comparators}

Human-comparator design, the extent to which human baselines constitute adequate benchmarks in capability evaluations, and how the lack thereof should be reported, is a first-order methodological question \citep{wei2025human}, which we draw on for the comparator-adequacy factor of the interpretive gap dimension. Suppose a capability claim in an AI capability evaluation paper is accompanied by a fully specified elicitation surface, but the task considered is one for which we would expect a human baseline, which the paper does not provide. Then, that capability claim is not fully interpretable because the paper's audience cannot situate the capability claim relative to an assumed professional baseline on the task in question. Human-comparator rigour is a necessary condition for interpretability of the capability claim, but not a sufficient one. The three audit dimensions jointly provide a sufficient condition for interpretability.

\subsection{The niche this paper fills}
\label{sec:background:niche}

This study addresses a specific gap in the four literatures above. Although Apollo Research, METR, and AISI identify the elicitation gap for model evaluations, they do not audit the manifestation of the elicitation gap in the academic literature. While CAIS's Safetywashing study evaluates a related but distinct concept (safety-claim inflation) using a similar methodology, their conceptualisation is not transferable to our study. Although existing bibliometric audits (Nagendran et al., Balloccu et al., Agrawal et al.) inform our methodology for the systematic coding of methods in a large number of studies, they do not break down the capability elicitation gap into its three fundamental mechanistic components (temporal lag, within-family tier lag, and configuration underreporting). Although existing reporting guidelines (CONSORT-AI, TRIPOD-LLM, DECIDE-AI, and STARD-AI) apply to related but distinct evaluations, none directly apply to the primary evaluation of interest (off-the-shelf capability evaluations). To our knowledge, this report represents the first pre-registered cross-domain measurement of the publication elicitation gap which decomposes it into three components (temporal, within-family tier, and configuration) and pairs it with a candidate reporting checklist (\versio{} v1.2) for authors, editors, and funding bodies to adopt in response to the diagnostic work presented herein.

\section{Methods}
\label{sec:methods}

\subsection{Pre-registration}
\label{sec:methods:prereg}

The protocol, as well as inclusion and exclusion criteria, were pre-registered on the Open Science Framework (\url{https://doi.org/10.17605/OSF.IO/7XM3D}). The pre-registration was time-stamped before the manuscript was deposited as a preprint on arXiv, and the analysis plan in this manuscript is the same as that in the pre-registration, except for the deviations in \S\ref{sec:lim:deviations}. Several primary magnitudes are purely descriptive and were pre-registered: configuration underreporting (H4), compound failure (H5), share of class-level claims, and the year-on-year slope (H2), which is reported standalone, with a journal-clustered bootstrap. Pre-registration also covers the \eci{}-gap outcome (including the missing-value imputation policy), the quaternary valence coding scale, the tier ladder defining the frontier, the sensitivity analysis for cross-family extraction, and the dual-human gold-standard validation design. Every falsification bucket in the framing maps is tied to a specific abstract commitment. All magnitudes except the H2 slope are bound under the framing maps.

\subsection{Corpus construction}
\label{sec:methods:corpus}

Our sampling frame is limited to OpenAlex. More specific databases exist for particular fields (e.g. PubMed/Embase for medicine, dblp for coding, and ERIC for education) that more thoroughly index their respective fields compared to OpenAlex. However, we chose to limit our sampling frame to OpenAlex because our first concern is comparability. We are testing our hypotheses (H1--H6) across five different domains and querying different databases for each domain would undermine this comparability.

We chose the five domains (medicine, coding, law, education, and scientific reasoning) because they are heterogeneous, i.e. different from each other. Much of the medical literature is focused on patient safety, and thus the field has many established guidelines for reporting AI studies (e.g. CONSORT-AI, SPIRIT-AI, TRIPOD-LLM, DECIDE-AI, and STARD-AI). In contrast, there can be wide variation in how coding tasks are configured, reflecting the heterodox methodological conventions of modern AI research. Somewhere between these two poles lie law, education, and scientific reasoning. In this way, comparing across these domains also serves as a test of whether the publication elicitation gap is field-specific or structural.

For a paper to be included in our sample, it must empirically evaluate one or more named LLMs on a task within an applied domain (medicine, law, coding, education, or scientific reasoning) and contain a quantitative result (e.g., accuracy, F1, BLEU, or another metric specific to the task). Further, it must have been published between 2022-01-01 and 2026-04-01, and be peer-reviewed or an arXiv, OSF, or SSRN preprint with an extractable link to the full paper.

We further exclude papers that only develop a prompting technique or similar without empirical evaluation in an applied domain, and papers that only evaluate in-house or unreleased models. Finally, we exclude duplicate papers (by DOI or by first author + year + primary benchmark).

We queried the March 2026 snapshot of OpenAlex \citep{openalex2024} for papers published between 2022-01-01 and 2026-04-01 that matched the search terms ``large language model'' OR ``LLM'' OR ``GPT'' OR ``ChatGPT'' OR ``Claude'' OR ``Gemini'' OR ``PaLM'' OR ``Llama'' OR ``Mistral''. Because of the year filter in the retrieval, 348 of the included papers ($1.9\%$) have a publication date in OpenAlex after 2026-04-01. We keep these papers in all analyses (the three dated after April 2026 fall outside the focal window); 345 of them have dates between 2 and 15 April, while two are dated in May and one in November 2026.

We first retrieved all results from OpenAlex for these search terms, then removed duplicate DOIs, papers not written in English, and grey literature (defined for our purposes as non-peer-reviewed work not published on arXiv, OSF, or SSRN). This retrieval was limited to 76,940 records by a row limit in the OpenAlex API. The initial retrieval (post-DOI deduplication, English-only, and peer-reviewed or preprint filter) contained 76,940 records (capped by OpenAlex API row limit). We then re-ran the exact same retrieval, but without the row limit. This identified 35,363 additional records that were dropped due to the cap (post-DOI deduplication, English-only, and peer-reviewed or preprint filter).

These two sets of records were combined to yield a set of 112,303 unique papers. We then re-ran the V4F classification system end-to-end on this set of papers.

Any papers that are not in the main five domains are placed in the ``other'' category, which we retain for descriptive purposes.

Each paper that evaluates more than one model contributes one paper-level record. The per-model dyads for H3 are in a separate file. The \texttt{primary\_model} for the record is the model with the highest \eci{} score among the extracted list of \texttt{models\_evaluated} and the extractor's own selection (which wins ties).

The full decision tree we use to determine whether a paper meets the inclusion criteria above is encoded directly in the frozen extraction prompt, which is available with the pre-registration on OSF.

\subsubsection{Coverage audit}
\label{sec:methods:coverage}

The corpus is not a census. Because title search using a set of keywords will fail to identify papers that evaluate LLMs but don't use any of those terms in their title or abstract (e.g., ``diagnostic accuracy of a conversational assistant for paediatric vignettes,'' which evaluated GPT-4 in the methods but didn't mention the model in visible text), we defined a ``residual pool'' of papers on the basis of two OpenAlex concept topics that encompass the audit scope: T11636 ``Natural language processing and large language models'' or T10181 ``Artificial intelligence in healthcare.'' We then applied a four-filter intersection to yield a pool of N=132,899 records: T11636 $\cup$ T10181 AND date $\geq$ 2023 AND work\_type $\in$ \{article, preprint\} AND not already in our integrated 112,303 paper corpus. We generated a random stratified sample of n=9,815 records from this pool and processed these using the V4F two-stage production pipeline (default-effort ai\_relevance classifier, max-effort v7.2 inclusion\_decision extractor). 436 of the records in the original random sample were incorporated into our integrated corpus as a result of the post-cap title-keyword expansion described in \S\ref{sec:methods:corpus}, leaving an effective n=9,379 record sample. Of these, 336 (3.58\%, Wilson 95\% CI: 3.22--3.98\%) were classified as inclusion\_decision = include. Extrapolating this sample-level inclusion rate to the population of the residual pool, we estimate the presence of approximately 4,761 (95\% CI: 4,286--5,287) additional LLM evaluation papers in this pool, suggesting a corpus capture rate of approximately 60\% (95\% CI: 57.4--62.5\%) among article/preprint records dated 2023 onwards on these two OpenAlex topics (7,138 included records in that window).

In a Bonferroni-corrected family of k=18 tests on the V4F-classified residual sample (n=336 inclusion\_decided), we find strong agreement between corpus and residual-pool outcome distributions. Beyond overall inclusion rate differences, we find a single significant compositional difference, which is an artefact of defining the boundary of the integrated corpus by title keyword search: in two primary model token cells, the residual sample overrepresents ChatGPT and underrepresents GPT-4. Neither Claude-3 ($p = 0.027$) nor differences in valence, framing, and the frontier gap proxy survive Bonferroni correction, though the proxy difference is significant at a nominal level ($p = 0.0043$), with the residual sample more distant from the frontier. See Appendix~G for details.

\subsection{Extraction pipeline}
\label{sec:methods:extraction}

We use a frontier LLM, rather than regular expressions or keyword matching, for this information extraction step because accurately reading the structure of a capability claim requires the same capability that is being claimed. There are two relevant peer-reviewed precedents for this approach. For inclusion, \citet{chen2026natmed} report Cohen's $\kappa$ = 0.820 for GPT-5 (high reasoning effort) on 500 gold-standard pairs from medicine-LLM papers. For extraction, \citet{ko2026jjr} report 85.9\%--100\% accuracy for GPT-4o and o1 on the objective MI-CLEAR-LLM items across 159 medicine-LLM papers. We pass the title and abstract of each paper to a frozen production prompt (ancillary file \texttt{appendix\_C\_\allowbreak frozen\_\allowbreak extraction\_\allowbreak prompt.md}) that elicits structured JSON outputs with associated confidence values for each field. We run this prompt on V4F (deepseek-v4-flash-max) at temperature 0.0 in a single pass with maximum reasoning effort (the highest reasoning-effort tier). We originally pre-registered the use of gpt-5.4-mini for both stages, but decided to switch to V4F for the entire pipeline due to cost and coverage considerations (the cost of V4F on the same prompt is $\sim$7\% of the cost of gpt-5.4-mini per token; see the deviation register in Section~\ref{sec:lim:deviations}). To verify V4F, we perform a four-extractor benchmark (Table~\ref{tab:benchmark-four-extractor}) and a cross-family LLM triad (Appendix~\ref{app:validation:crossfamily}). We find that the v4f$\leftrightarrow$opus LLM-extractor pair, in particular, achieves $\kappa$ $\geq$ 0.65 on every load-bearing subjective field.

For the $n = 4{,}766$ papers with a retrievable, machine-readable PDF, we use two other hardened prompts to extract additional information from the full text (per-pass extraction success differs by field: $4{,}757$ records returned a usable eval-date pass, $4{,}762$ a usable elicitation-field pass). The first such prompt attempts to extract the \texttt{evaluation\_date} and the description of the primary model. The prompt explicitly forbids using various types of proxy information (submission/acceptance/publication dates, copyright year, model training cutoff or release dates, benchmark publication and dataset collection dates, or dates for other past studies) for the evaluation date. We used a deterministic resolver to match model names with canonical models and release dates, and assigned bare family names (e.g., ``GPT-4'') to the initial release date for that model. The resolver routes based on the release date (distinguishing between pre- and post-March 2023 versions of ChatGPT), and passes through specific snapshot names. The prompt used to elicit this information, and the table of model release dates (sourced from Epoch's database), are both available in the data bundle associated with this paper. The canonical model resolver is available in the code deposit associated with this paper (see Data availability). SHA-256 hashes of the companion prompts are listed in the manifest in \texttt{appendix\_C\_\allowbreak frozen\_\allowbreak extraction\_\allowbreak prompt.md}. The second companion prompt was used to extract the six fields related to elicitation configuration used for H4 and H5. Each field is gated on applicability flags set for each (\texttt{model\_surface}, capability) pair.

For the full-text statistics, we condition on the full-text pass's own primary model (4,753 records), with the 9 records lacking such a model defaulting to the abstract-level primary model. Since the full-text pass has no record of the list of evaluated models (we checked all 4,766 full-text records), the registered highest-\eci{} rule cannot be applied to it, and the abstract-level and full-text primary models are differently defined under that rule. The rule affects the abstract-level primary model of 506 full-text papers. For 130 of these, the model selected by the rule was already present in the full-text pass. For the other 376, the use of the abstract-level primary model results in an overall disclosure rate for reasoning mode of 21.1\% (108/512), which we compare to the reported rate of 21.2\% (111/524).

We provided the extraction model with the title and abstract of each paper and had it make an inclusion decision, as well as extract the domain, models evaluated, primary model, and whether the paper reported an evaluation date, along with the eight configuration fields (access method, temperature, prompting strategy, reasoning mode, thinking effort, tool use, scaffolding, and multi-agent architecture). We also had it identify whether the conclusion of the paper was about the model being tested, or about language models as a class.

We did not have the resources to run a human coding panel of this size ($n = 18{,}574$) and matched to model tier (see Section~\ref{sec:lim:pipeline}). Therefore, we checked convergent validity at corpus-scale using a cross-family triad ($n = 150$). See Appendix~\ref{app:validation:crossfamily} for the sample, models, and agreement floors for this triad, as well as for a second triad for convergent validity (again, $n = 150$) in which V4F replaced one pre-registered model. In a robustness check (Section~\ref{sec:methods:sensitivity}), we run a sensitivity analysis where we drop all papers with extraction confidence below 0.90 and re-analyse the data. We will share confidence flags for all papers with any researcher upon request, so that researchers can filter the data to their desired level of confidence and re-run analyses.

\subsection{Validation protocol}
\label{sec:methods:validation}

We sampled the gold standard from an oversample of the initial gpt-5.4-mini classification of the entire corpus. To validate against the pre-registered production extractor on both the subjective fields and the inclusion decision, we re-extracted the entire oversample using the cross-family triad from our pre-registration. The subjective fields were independently coded by two blinded coders, D.G. and M.S., and disagreements were adjudicated by an LLM applying the coding rubric under D.G.'s direction. Table~\ref{tab:kappa-7-2} shows the samples that were used for this analysis and the pre-registered $\kappa$ floors and $\kappa$ values for each field.

After coding disagreements were adjudicated, the consensus-set papers were used as a reference standard to score the extraction model. The class-level claim share (Sections 3.4 and 4.2.7) is estimated using a Bayes-corrected estimator that imputes the \texttt{conclusion\_framing} indicator using the gold confusion matrix. The gold confusion matrix was computed on the merged (post-adjudication) dual-coder consensus set ($n = 231$). This estimator is intended to correct defensively for residual error in the production extractor with respect to the gold standard. We also report the gold-anchored direct count on this subset (53.2\%) as a non-parametric anchor.

For H6, we simulate measurement error by drawing 1,000 times from the empirical distribution of misclassification. We then claim a rejection only if at least 90\% of the simulations show the hypothesised direction of the effect.

\subsection{Frontier measure}
\label{sec:methods:frontier}

To contextualise AI capabilities during each study's evaluation period, we leverage the Epoch AI Capabilities Index (\eci{}) to establish the state of the art on each evaluation date. Data on the capabilities of frontier models were gathered from Epoch AI's April-2026 capabilities snapshot \citep{epochai2024} (data/eci\_scores.csv), which contains the calibration data on \textasciitilde{}165 frontier or near-frontier models across 1,470 model/benchmark cells. \eci{} is anchored to an AI capability score of 150 for GPT-5, released in August 2025. We define the AI capability frontier as the highest-\eci{} model publicly or commercially available as of each study's evaluation date. Our \eci{} gap measure, therefore, is the difference between the \eci{} of this model and that of the primary model used in a given paper. When a specific evaluation date is not able to be extracted from full text, following our pre-registration, we impute it as the maximum of publication date minus 180 days and the model release date, if tabulated (see Appendix~\ref{app:spec} for a sensitivity analysis to this choice). In cases where we extract an evaluation date from paper full text, this is used in lieu of our imputed date. Of 4,757 papers with successfully extracted full text, 877 mention an evaluation date in their methods section. Of these 877 papers, 872 give a parseable date, and are therefore used in our ``lag default'' sensitivity analysis. The remaining five papers disclose an unparseable evaluation date (these are discarded from the imputation-override layer, but included in our raw eval date disclosure rate). Our default of 180-day cross-domain lag approximates the median lag between submission and publication in the corpus, weighted by domain. We also test alternative defaults of 0 days, 90 days, 270 days, and 365 days, as well as the median within each domain (see Table~\ref{tab:lag-default-sensitivity} for all results). When a canonical model has no score in the \eci{} database, the paper is excluded (11 papers); no score is imputed from a sibling model. Our analysis sample for H1 and H2 consists of the n=12,668 papers in our imputed-anchor analysis sample. For H2 and H3, we conduct separate filtering steps, resulting in a sample size of n=12,242 for the journal-clustered analysis, and n=4,014 papers for which we can conduct a within-family sibling look-up analysis. In the latter analysis, we represent each paper by its primary model and the qualifying sibling model with the highest \eci{}.

We conduct a sensitivity analysis in which we read the registered frontier steps on the first day of every month. This means that the frontier has a month-start granularity. The only change is to the time granularity, and the frontier window opens on 2023-03-01. The results are reported in Appendix~\ref{app:spec}. We also operationalise a deployment capabilities frontier as a sensitivity measure: for each research evaluation date, we define the deployment frontier as the model with the highest \eci{} among those models with an API price per token of at most 10x that of the cheapest non-frontier (``base'') tier model available on that date. Given that prices of frontier-tier models tend to vary by less than an order of magnitude at any given time, this operationalisation of the deployment frontier focuses on the bottom of the pricing distribution (as opposed to the top) to avoid trivially including all frontier models. This deployment frontier measure is anchored to prevailing market pricing at each research evaluation date and thus co-varies over time with model pricing. The deployment frontier measure is specified for the exploratory H10 invariance test, which is not reported here.

We use \eci{} as our primary frontier model capability measure because the next best alternative measure, the calendar recency in terms of the number of months between model release date and evaluation date, is not able to capture differences in model capability between models of different capability tiers. As capability tier differences are an important feature of the models used for research evaluations in our corpus (e.g., Claude 3 Opus and Claude 3 Sonnet were released on the same day but have meaningfully different \eci{} scores), a measure of frontier distance based solely on calendar recency would not be able to differentiate between these models (in contrast to \eci{}). The protocol also specifies a tertiary domain-specific frontier gap (the highest-\eci{} model evaluated in a given domain corpus year) and a per-benchmark-cluster frontier gap based on task-matched Epoch benchmarks; neither is reported here.

As with any scalar measure of capabilities, there is no way to fully represent the multidimensional capabilities profile without losing some information. As noted in Epoch's methodology, models specialised on narrow domains ``may receive low \eci{} scores, despite being very capable within their domain'' \citep{epochai_eci_methodology}. Epoch also points out that developers ``can optimize for high performance on certain benchmarks.'' Epoch mitigates the risk of benchmark overfitting by conducting its own internal evaluations, as well as drawing on evaluations from independent leaderboards. Still, training-time awareness of these benchmarks and frontier-scale compression remain important limitations of every benchmark-aggregated index.

We view the use of \eci{} as our primary scalar summary of frontier-ness as the least-bad option, compared to the other audit-relevant options, which all fail in at least some load-bearing cases. Using calendar-time recency as a scalar fails to discriminate between the capabilities tiers discussed above. Using a per-benchmark matching methodology will fail to yield an assessment in applied-domain tasks that have a tenuous mapping to the benchmarks used in Epoch's \eci{} methodology. And qualitative expert rankings of models re-introduce the kinds of researcher degrees-of-freedom we want to avoid in this audit. For these reasons, we use \eci{} as the primary scalar summary of frontier-ness in the audit. We are cognizant of the important limitations and caveats that Epoch highlights about this metric and take those into account throughout the audit. As Epoch writes, ``absolute \eci{} values are meaningless by themselves, but meaningful comparisons can be made between models'' \citep{epochai_eci_methodology}. In the audit, we exclusively interpret pairwise \eci{} gaps, rather than absolute \eci{} values.

We also view calendar-time recency as a companion measure to \eci{}, and do not make any primary interpretations of \eci{} gaps unless they accord in sign with calendar-time gaps. We also report the three constituent components (temporal, capabilities tier, and configuration) as a vector to enable re-weighting.

Finally, we present results using the Artificial Analysis intelligence index (AA), an alternative scale of model capability that independently aggregates multiple benchmarks into a single index. We provide median estimates for H1, slope estimates for H2, and conditional magnitude estimates for H3 for each of \eci{}, Arena Elo, and AA. To produce the scale sensitivities for the class-level framing trend, we perform domain-adjusted regressions using only the subset of papers covered by each scale. In these regressions, no capability score is included as a covariate. Neither the confirmatory sign tests nor the framing-map bucket assignments are required to hold for all scales. By explicitly reporting the dependencies of various claims on alternative capability scales, we enable readers to identify which claims are contingent on any particular capability scale. Appendix~\ref{app:eci} reviews the construct validity of \eci{} as a proxy for frontier-ness for the purposes of this audit, and reports the Pearson correlation between \eci{} and Chatbot Arena Elo scores, for models scored by both scales.

\subsection{Primary outcomes and hypothesis tests}
\label{sec:methods:outcomes}

We report results using measures at three levels of granularity. At none of these levels do we combine our audit dimensions into a weighted composite. First, the headline measure we report is the compound failure rate (CFR): the proportion of included papers that fail on all three audit dimensions. The CFR is a primary descriptive proportion measure. Its value depends on the exact definition of failure on each audit dimension. A paper fails on the capability dimension if it has \texttt{eci\_gap} $\geq$ 12.0 \eci{}. This is based on the \eci{} jump between models one major generation apart, averaged over pairs of models in the same family and tier, using \eci{} data from Epoch's frozen April 2026 snapshot. The pairs, with corresponding \eci{} jumps in brackets, are: Claude 3.5 Sonnet vs Claude 3 Sonnet [9.92], Claude Opus 4.6 vs Claude Opus 4 [11.72], Claude Opus 4 vs Claude 3 Opus [16.30], and Gemini 1.5 Pro February 2024 vs Gemini 1.0 Pro [11.22]. The average of these four values is 12.29, which we round to 12.0. A paper fails on the elicitation dimension if it meets any of three conditions (the primary ``OR-of-three'' failure definition): the paper ran its evaluation zero-shot or at default effort, even though at the time of the evaluation, the family of the model being evaluated had a scaffolded baseline; tested a reasoning-capable model, but did not report whether reasoning was enabled; or tested a tool-capable model, but did not report whether that model was given tools. A paper fails on the interpretive dimension only if it both lacks a human or professional comparator (after applying the domain-based admissibility rule) and has \texttt{conclusion\_framing} = \texttt{ai\_generic} (the ``AND-of-two'' operationalisation). For two of our dimensions (capability and interpretive), the bar for failure using our primary definition is conservative. The CFR is therefore a descriptive and conservatively biased estimator of the proportion of included papers that have a latent compound failure, rather than a point estimate of that proportion.

Second, we use as a secondary measure the capability elicitation shortfall: \texttt{eci\_gap}*(1 - \texttt{config\_elicitation\_index}). This is intended to capture the interaction between the capability gap and the degree of elicitation disclosure. Using the gap or disclosure measures alone would not capture this. For instance, a paper that disables reasoning when evaluating GPT-5.5 Pro would fall short differently from a paper that uses well-scaffolded elicitation for a GPT-4o evaluation. The configuration elicitation index is a simple mean of six model configuration components (reasoning mode, thinking effort, tool use, scaffolding, multi-agent architecture, and prompting strategy), weighted equally. Components that are not applicable are omitted. We use a published admissibility list to make judgement calls about edge cases, like the interaction between task type and tool use. For transparency, we also report a tertiary measure: a vector of (\texttt{temporal\_gap}, \texttt{tier\_gap}, \texttt{elicitation\_gap}), where the entries always appear in this order. Section~\ref{sec:results:confirmatory} presents the overall magnitudes of these vector components in the corpus as primary measures. We report the pooled \texttt{eci\_gap} (H1): for dyad-eligible papers, this is the sum of the temporal and tier gaps, and for other papers, it is equal to the temporal gap. We also report the within-family tier gap for dyad-eligible papers (H3) and the elicitation disclosure gap for the applicability-conditioned subset (H4). We do not include a separate, redundant decomposition table for the corpus overall. Readers may wish to weight the component gaps differently; as we always report the vector, readers can combine these components in different ways. The data bundle (provided by the authors upon request) contains the inputs to the vector components for each paper.

We have three confirmatory hypotheses, H1, H3, and H6. These hypotheses share a directional-confirmatory testing strategy with a Holm correction, carried out as a step-down procedure at a family-wise $\alpha$ = 0.05. Although we use ``directional sign-test'' as a catch-all term for this general approach, this is not a reference to the statistical sign-test procedure. For all null hypotheses, we are dealing with structural zeros and do not introduce any rejection thresholds of our own choosing. Both H1 (location) and H3 (tier lag) are implemented as one-sample Wilcoxon signed-rank tests. The null hypothesis for H1 is that the median \texttt{eci\_gap} does not exceed zero. It is rejected if and only if the median's point estimate is greater than zero and the Holm-adjusted p-value is below $\alpha$. The null hypothesis for H3, which is based on the \texttt{tier\_gap} variable (with the number of eligible papers as the denominator), defined as \eci{}(\texttt{best\_sibling\_in\_window}) -- \eci{}(\texttt{tested\_model}), is that the median tier gap does not exceed zero. A ``qualifying sibling'' is one that is part of the same model family, has a higher \eci{}, was released by the time of the paper's evaluation, and was released within 90 days of the release of the tested model. Given the above eligibility requirements, the gap is guaranteed to be positive on the primary scale. The H3 distribution thus reflects the conditional magnitude of the gap. Therefore, it cannot be used to demonstrate the presence of the effect by itself. For H6 (valence asymmetry), we implement \texttt{eci\_gap} \textasciitilde{} \texttt{conclusion\_valence} + domain + year + domain:year + (1|journal) as a mixed-effects model and test a directional contrast within it. The beta coefficient for valence=negative vs. valence=positive is the contrast of primary interest. Two pre-registered fixed-effect covariates, \texttt{author\_affiliation\_type} and \texttt{venue\_type}, were ultimately not included in the model (see Deviation Register; \S\ref{sec:lim:deviations}). The H6 null hypothesis is rejected if the beta coefficient for valence is greater than zero, its two-sided 95\% confidence interval does not include zero, and its p-value is below $\alpha$ after Holm adjustment. Beyond this rule, an additional condition must be met: the sign must be robust to measurement error in the simulation described in Section~\ref{sec:methods:validation} in order to reject the null. We define this as the association preserving its sign in $\geq$90\% of 1,000 OLS-HC3 draws under the measurement-error model. As a sensitivity analysis for H6, valence is given a numeric linear coding. Before the Holm adjustment, the one-sided p-values from H1 and H3 (each doubled, upper-bounded at 1) are combined with the two-sided p-value from H6.

Of the four primary descriptive magnitudes that we report, three are part of a descriptive family which we subject to Holm-Bonferroni, while the fourth, H2, is not part of that family. H2 is the OLS slope in \texttt{eci\_gap} $\sim$ \texttt{publication\_year} + domain + domain:year (clustered at journal), or $\hat{\beta}$, and the abstract claim has a pre-registered directional-sign test that could falsify the thesis ($\hat{\beta}$ < 0 with a confidence interval entirely less than 0 would be evidence against the part of the thesis about persistence). We report inference on H2 on its own (providing a confidence interval using a journal-cluster bootstrap) because it is a directional-sign falsifier and not a level threshold. H4 is the rate of \texttt{reasoning\_mode} disclosure among papers that test reasoning-capable models. H5, the compound failure rate, uses conservative definitions for its capability and interpretive components, and as a descriptive estimator of the latent compound failure share is conservatively biased. Whereas H1, H3, and H6, the confirmatory tests, all use directional rules and a structural zero null, H5 is different in that it is a reported proportion under a pre-registered conjunction. The class-level claim share is the proportion of included papers whose abstract is coded \texttt{ai\_generic} on \texttt{conclusion\_framing}. We report the class-level claim share using our marginal posterior per paper (see \S\ref{sec:methods:validation}) as our primary estimator of this proportion, and report both the level and the trend over publication years of this share. We report Holm-Bonferroni simultaneous 95\% confidence intervals for a three-member family consisting of the class-level claim share, H4, and H5. Our pre-registration framing map (preregistration/\texttt{framing\_maps}.md) links all members of this family to falsification buckets, and those falsification buckets have falsification values H4 $\geq$ 0.50, H5 < 0.02, and class-level claim share < 0.05.

We report the results of our exploratory tests (H7--H9) without claims about $\alpha$-levels. H7 is a test of dispersion. We test for bimodality in \texttt{eci\_gap} using Hartigan's dip test. H8 uses delay before publication to contextualise the capability gap. However, external benchmarks for delay before publication cannot by themselves identify the degree to which publication delay is responsible for the capability gap. H9 disaggregates the H1--H3 estimates by domain. H10, where we would repeat the confirmatory sign tests for H1, H3, and H6 and compute the descriptive estimates for H2, H4, and H5, as well as the class-level claim share, after substituting in the deployment frontier, is not reported.

\subsection{Sensitivity analyses}
\label{sec:methods:sensitivity}

We pre-registered a number of sensitivity analyses: subsetting the descriptive and confirmatory analyses by domain, re-running them by publication year, comparing journal with arXiv/conference papers, and dropping all papers with an extraction confidence score below 0.90. For papers that benchmark multiple models, we also pre-registered a dyad-level analysis and within-paper fixed effects for H6. We did not execute the within-paper fixed effects for H6 because, with a single label for \texttt{conclusion\_valence} at the paper level, the variable of interest would be absorbed by the fixed effect, and the protocol did not suggest a within-paper contrast that would identify H6. Pre-registered alternative definitions include an AND-of-three version of elicitation failure (rather than the primary OR-of-three) and an OR-of-two version of interpretive failure (rather than the primary AND-of-two), as well as the binary valence fallback definition and, for H5, a capability threshold at the 50th, 75th, 90th, and 95th percentiles of the empirical distribution of \texttt{eci\_gap}. The registered primary treatment of bare, unversioned mentions of ``ChatGPT'' is to impute GPT-3.5 Turbo (June 2023), the earliest \eci{}-scored GPT-3.5 model. Our main sensitivity analysis instead uses date-aware labels for bare mentions of ``ChatGPT'': pre-2024-05-13, this is taken to refer to GPT-3.5, and post-2024-05-13, it is taken to refer to GPT-4o; and in the context of a paper that mentions ``Plus'', it is taken to refer to GPT-4 from 2023-03-14. This analysis yields an H1 median of 9.01 [8.36, 9.73] ($n = 12{,}685$). Table~\ref{tab:registered-sensitivity} reports the other two bare-ChatGPT treatments (anti-hypothesis and exclusion) as well as the results of other pre-registered subset and alternative definitions. We examine the full-texts of bare-ChatGPT papers in Appendix~\ref{app:spec:registered}.

In H10, a pre-registered invariance test, the factor-10 deployment-accessible capability frontier is substituted in for the primary capability frontier. This test would examine the stability of descriptive statistics (H2, H4, H5, and the class-level claim share) and confirmatory signs (H1, H3, and H6) under this substitution. We did not execute H10 because the deployment-accessible capability frontier requires an API price for each model, on each evaluation date, over 2022--2026, and we only have notes on the current models' extraction cost in April 2026.

Our final robustness test examines the consequences of not imputing missing \eci{} scores. No \eci{} score in the analysis sample is imputed: every H1 primary model has its own \eci{} entry, and the 11 papers whose canonical model has none are excluded. H1--H3 as reported are therefore already estimated without imputed \eci{} observations.

\subsection{Measurement map}
\label{sec:methods:measurement-map}

The ancillary file \texttt{measurement\_map.md} lists the primary and secondary outcomes of the audit, and for each outcome, the paper surface from which it is read, the extraction artefact it depends on, its analytic n, and whether it was pre-registered.

\section{Results}
\label{sec:results}

\subsection{Descriptive findings}
\label{sec:results:descriptive}

\begin{figure}[!htb]
  \centering
  \includegraphics[width=0.95\textwidth]{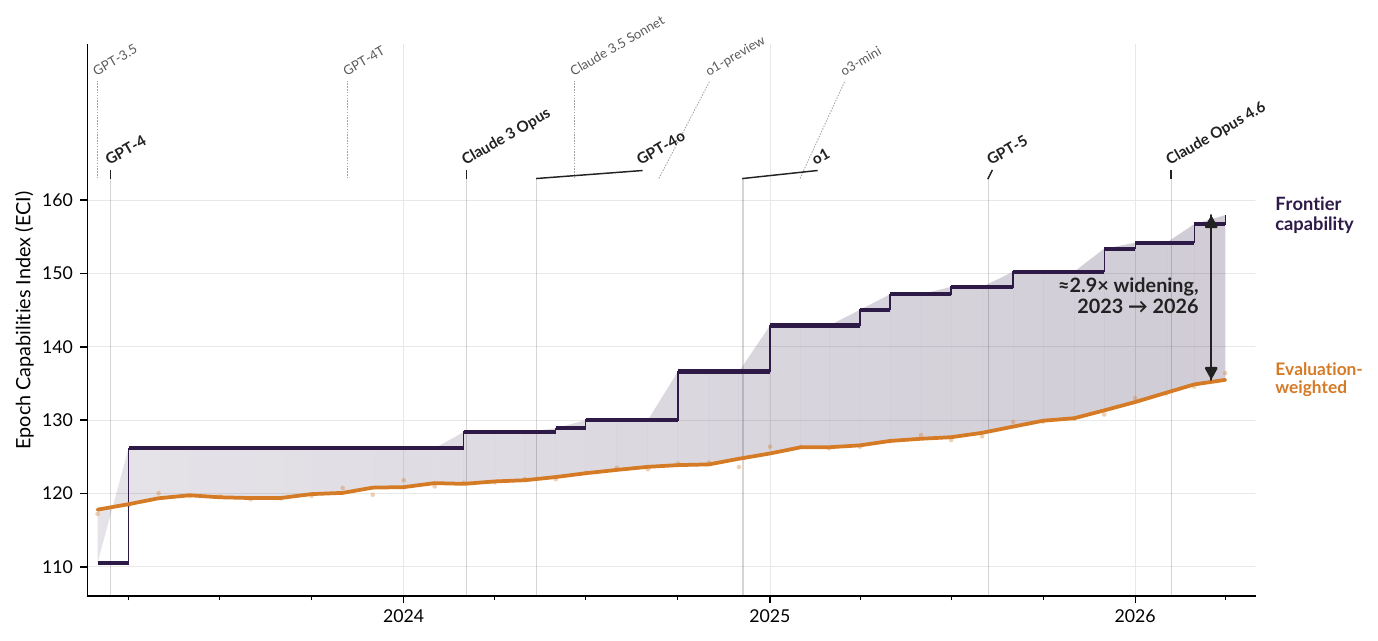}
  \caption{Registered on-date frontier \eci{}, read at month start (stepped; key frontier-model release dates are annotated). In burnt orange, evaluation-weighted, centred three-month rolling mean of published-paper primary-model \eci{} (highest-\eci{} model listed; registered rule) by publication month, 2023-03 to 2026-04, V4F-cascaded full corpus, $n = 12{,}526$ \eci{}-resolvable papers. Shading marks the separation. The $\approx$2.9$\times$ widening label is 22.46 \eci{} (2026-04 gap) over 7.64 \eci{} (first positive gap, 2023-04).}
  \label{fig:trajectory}
\end{figure}

We refer to three different corpus-level $n$ values throughout our analysis: $n = 18{,}574$, representing the total number of papers included; $n = 18{,}565$, representing the number of V4F-cascaded papers for which we were able to extract a \texttt{conclusion\_framing}, used as the denominator for class-level claim shares; and $n = 18{,}311$, representing the number of papers in our focal period from 2023-03 to 2026-04, which excludes papers published prior to 2023-03, whose contemporaneous frontier would lie before our H2 fit window. Figure~\ref{fig:trajectory} focuses on this period. Specific outcomes have their own, smaller $n$ values due to outcome-specific filtering, which are documented in the ancillary file \texttt{measurement\_map.md}.

The production pipeline processed $n = 112{,}303$ papers from the OpenAlex pool. Of these, $n = 18{,}574$ papers met the pre-registered admissibility criteria and were included in the final dataset. Most papers are in medicine or the residual category of cross-disciplinary research. The remaining four domains each make up single-digit to low-double-digit percentages of the included papers.

Under our pre-registered rule, the primary model of a paper is the highest-\eci{} model that the paper evaluates. In 810 of 8,775 papers from 2025 and 185 of 3,868 papers from 2026, the primary model is still GPT-3.5 or an even earlier model. Of these papers, 174 (2025) and 33 (2026) explicitly use the names GPT-3.5 or GPT-3. The majority of the remaining papers do not specify the version of ``ChatGPT'' they use, which our resolver maps to GPT-3.5 Turbo (June 2023). Primary models have a heavy-tailed distribution. Four model families make up 86.7\% (11,026/12,720) of papers with a primary model that resolves to a known model: OpenAI GPT (excluding OpenAI's o-series models), Anthropic Claude, Google Gemini, and Meta Llama. GPT-3.5 Turbo is the most common primary model in every quarter of 2023, GPT-4 is the most common primary model from 2024Q1 to 2025Q3, and GPT-4o is the most common primary model in 2025Q4 and 2026Q1. By year, the most and second-most common primary models are: GPT-3.5 Turbo (713 papers) and GPT-4 (581 papers) in 2023, GPT-4 (1,354) and GPT-3.5 Turbo (894) in 2024, GPT-4o (1,199) and GPT-4 (1,148) in 2025, and GPT-4o (258) and GPT-4 (206) in 2026 (a partial year).

Splitting out configuration-reporting rates by cohort of paper publication (pre-2023, 2023, 2024, 2025, 2026) reveals the core structural pattern our audit is designed to detect. Temperature and sampling settings are reported consistently, and the first reasoning-mode dials were introduced with o1 in late 2024. Reasoning-mode status is not reported in earlier cohorts and is reported much more often in 2025 and 2026. Scaffolding is reported quite rarely, as is the use of a multi-agent architecture. Reporting norms have adapted slowly to changes in the range of possible configurations.

Figure~\ref{fig:trajectory} is our headline figure. The frontier is shown as a step function for the \eci{} of each month, with major model releases labelled along the line. We also show the \eci{} of the primary model used in each paper, as reported in the paper, smoothed using a centred 3-month rolling mean. We compute this over the full V4F-cascaded corpus for all papers between 2023-03 and 2026-04 for which we could resolve the primary model \eci{} ($n = 12{,}526$ out of 18,311, 68.4\%). Typically, the frontier \eci{} is shown on top and the primary model \eci{} on the bottom, with the area between shaded to indicate the gap. We see that this gap expands by a factor of $\sim$2.9 from 2023 to 2026. From mid-2023 on, the frontier \eci{} is higher than what is reported in papers. For a brief period at the left-hand side of the panel, the two lines intersect.

\subsection{Confirmatory hypothesis tests}
\label{sec:results:confirmatory}

The three anchor outcomes correspond to the three audit dimensions. The remaining hypotheses (widening trend, H2; within-family tier lag, H3; compound failure rate, H5; valence asymmetry, H6) decompose the anchor findings and stress-test them. The measurement error simulation for H6 draws on the gold confusion matrix. We run the 3 confirmatory directional tests (H1, H3, H6) under Holm step-down at family-wise $\alpha$ = 0.05. See \texttt{measurement\_map.md}, an ancillary file mapping all six hypotheses (descriptive and confirmatory) and the class-level claim share to their roles in the thesis argument.

\subsubsection{H1 -- location of the gap}
\label{sec:results:h1}

We find a gap of 10.45 \eci{} points (IQR: 1.12 to 16.72; 95\% bootstrap CI: 10.31 to 10.45; $n = 12{,}668$). This estimate is based on the pre-registered imputation strategy, whereby we assume evaluations occurred 180 days prior to publication if no full-text evaluation date is provided (see Section~\ref{sec:methods:frontier}). Under this imputation strategy, a one-sided Wilcoxon signed-rank test is highly statistically significant (p<$10^{-300}$; at double precision, SciPy returns p=0; Holm-adjusted p-value is also too small to represent). A one-sided sign test is also highly statistically significant (gaps: 10,341 positive, 153 negative, and 2,174 zero; p<$10^{-300}$). This test assesses the median gap and does not rely on the same symmetry assumptions. Thus, we reject the null hypothesis of a structural zero gap at post-Holm-adjusted $\alpha$=0.05, and confirm the sign of H1. For the median paper, the evaluated model is almost one model generation behind the frontier as of the paper's evaluation date. This gap is about 1.3 times the gap between Claude 3.7 Sonnet and Claude Opus 4.5, a step within a single model family that involves a major version change and a ``product tier'' change.

We also conduct the analysis on a sub-corpus of papers ($n = 746$) that only includes \eci{}-resolvable papers with the evaluation date disclosed in the methods section of the paper. In this sub-corpus, we do not perform any imputation. We find that the median value of H1 is +3.52 \eci{} (IQR [0.00, 12.63], p < $10^{-78}$, Wilcoxon one-sided test), again rejecting the structural-zero null hypothesis.\footnote{Of the $872$ disclosed-date papers, $88$ have no canonical model with an \eci{} score and $38$ disclose an evaluation date before March 2023, where the monthly frontier series starts ($32$ of them resolve to GPT-3.5 Turbo, June 2023); both groups are excluded.}

We also perform this analysis on each domain independently (Appendix~\ref{app:spec}). We find that we can reject the structural-zero null hypothesis for every pre-registered domain (all p < $10^{-21}$, Wilcoxon one-sided test). The median value of H1 varies between domains, ranging from +3.09 \eci{} (scientific reasoning) to +12.63 \eci{} (education). In no domain is the sign of H1 reversed.

We also conduct a sensitivity analysis (Appendix~\ref{app:spec}, Table~\ref{tab:lag-default-sensitivity}) wherein we vary the imputation lag (the number of days between the paper's publication date and the imputed anchor date) across \{0, 90, 180, 270, 365\} days, as well as a variant where we use domain-specific median lags. We conduct this analysis on all three independent capability scales. In all cases, the sign of H1 is preserved, though the magnitude varies. The median value of H1 in the pooled analysis ranges from +3.83 \eci{} (365 day lag) to +14.84 \eci{} (no lag). This analysis confirms the sign of our estimate of H1 using our imputed anchor date, though not the magnitude.

The sign of H1 also stays the same whichever definition of primary model in Table~\ref{tab:primary-model-definitions} of Appendix~\ref{app:spec:primary-model} is applied. The median value of H1 across the shortest and longest lag bounds for each paper ranges from 10.45 to 14.01 \eci{}.

\subsubsection{H2 -- trend over time}
\label{sec:results:h2}

The gap is not shrinking. We fit the pre-registered model with \texttt{publication\_year} as a predictor of \texttt{eci\_gap}, including fixed effects for domain, interactions with domain, and journal-clustered standard errors. This generates a slope estimate for each domain; we pool these slopes, weighting by number of papers. We find that the gap is increasing with a pooled slope of $\hat{\beta}$ = +4.07 \eci{}/year (95\% CI [+3.75, +4.45] bootstrapped by resampling journal clusters; $n = 12{,}242$ papers in 2,361 journal clusters). This positive slope was observed in every pre-registered domain.

The field is renewing itself more slowly than the frontier is being released. We pre-registered a directional-sign falsifier for the thesis: $\hat{\beta}$ < 0 with a confidence interval excluding 0 on the negative side. We do not trigger this falsifier. In every pairing of imputed lag and capability scale, the median distance and its year-on-year growth remained positive (Table~\ref{tab:lag-default-sensitivity}).

We consider several possible explanations for this widening gap, including under-reporting of the elicitation surface, publication lag, and API access limited by cost, but we do not attempt to estimate the relative contribution of each.

\begin{figure}[!htb]
  \centering
  \includegraphics[width=0.95\textwidth]{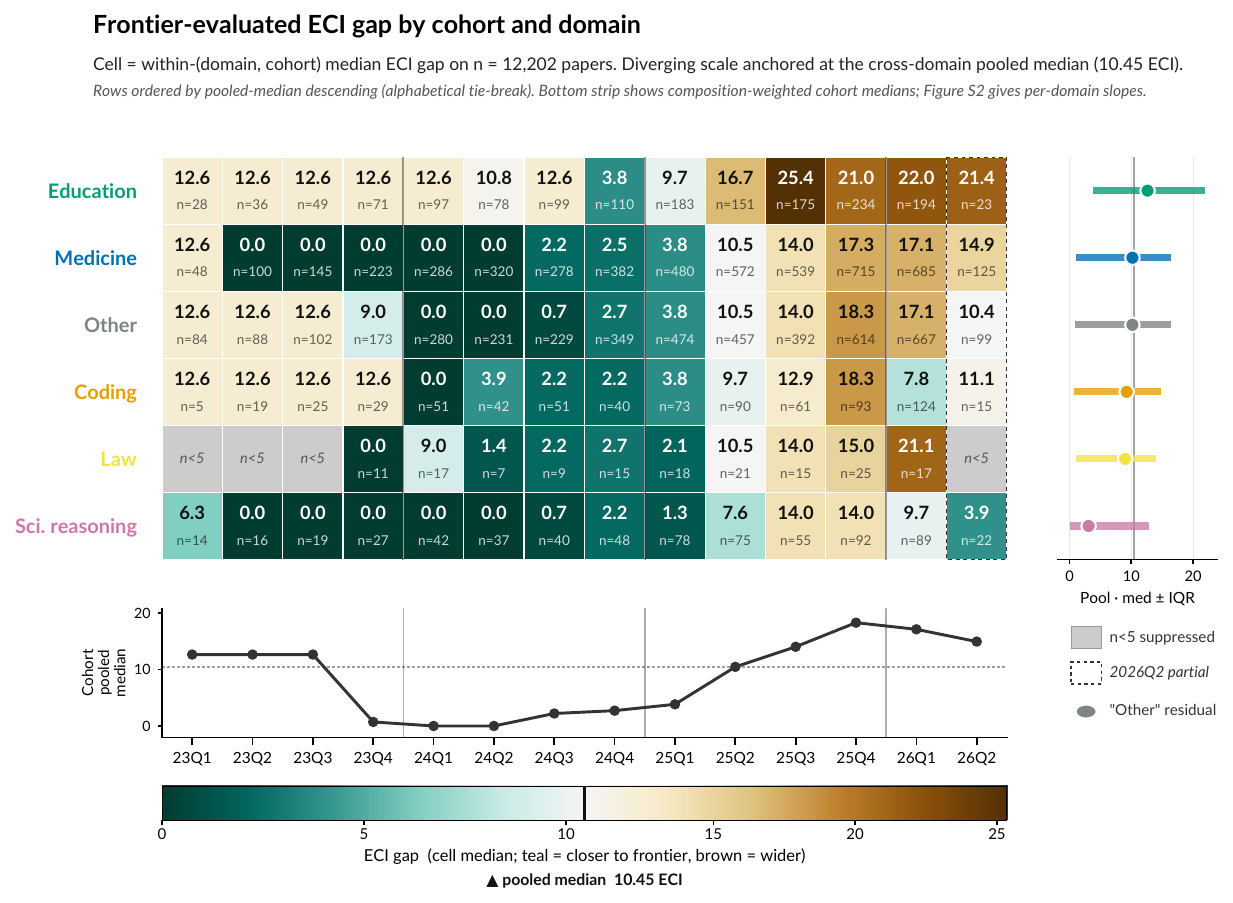}
  \caption{Median \texttt{eci\_gap} for each (domain, cohort) cell, with marginal distributions for each domain (pooled median and IQR) and cohort (pooled median over time, bottom). Diverging colour scale is anchored at the cohort-windowed pooled median ($+10.45$ \eci{}). Cells with $n < 5$ are suppressed. The fourteen cohorts are aggregated by publication date quarter. \texttt{eci\_gap} is computed at evaluation date. Sample: Cohort-windowed, analysable subset ($n = 12{,}202$) of the full dataset described in \S\ref{sec:methods:frontier} ($n = 12{,}668$ with 180-day imputation). The difference is 426 papers without a record in the corpus and 40 more papers published outside the 2023Q1--2026Q2 window. Five pre-registered domains, plus the residual \texttt{other} category, are shown.}
  \label{fig:domain-cohort-heatmap}
\end{figure}

In Figure~\ref{fig:domain-cohort-heatmap} we decompose the widening gap (H2) across domains (the five pre-registered domain partitions plus an ``other'' residual) and over time (fourteen quarterly cohorts of papers). The cells are coloured using the BrBG diverging scale, centred on the pooled median of that cohort window, which is +10.45 \eci{} (compared to +10.45 in the full-sample H1 analysis), with teal indicating cells closer to the frontier and brown indicating cells further away. The right panel shows the marginals for rows, pooling across cohorts within a domain, and the bottom strip shows the cohort-pooled median, weighted compositionally across domains (2026Q2 is a partial quarter). We observe a non-monotonic dip in 2024.

\subsubsection{H3 -- tier lag}
\label{sec:results:h3}

For H3, we take all papers where the evaluated model has at least one higher-\eci{} sibling model in its model family; the sibling model must have been released by the time of evaluation, and be released within 90 days of the evaluated model's release. This subset, using the pre-registered imputation strategy, is $n = 4{,}014$ dyad-eligible papers. In this subset, the median \texttt{tier\_gap} is +12.63 \eci{} (one-sided Wilcoxon p-value<$10^{-300}$; output from SciPy is a p-value of 0 at double precision). The Holm-corrected p-value for this hypothesis, within our family of confirmatory hypothesis tests, is also less than the minimum representable value at double precision. We thus reject the associated null hypothesis of structural zeros. Of the 4,014 H3 papers, 2,588 pair GPT-3.5 Turbo (June 2023) with GPT-4 (March 2023), whose snapshot dates are exactly 90 days apart. OpenAI dates GPT-4's release to 14 March 2023 (91 days), one day outside the registered $\pm$90-day window. If we move only GPT-4's date, the H3 median becomes +8.54 \eci{} ($n = 3{,}965$), and the test still rejects the null of a structural zero. Because this hypothesis only considers eligible dyads with positive \eci{} gaps, the sign of the difference is predetermined. Therefore, though we can reject this structural zero null, we cannot use this test alone to confirm the presence of an effect. However, this dyad-eligible subset allows us to characterise the magnitude of the tier lag, conditioning on the subset of papers with coverage of a non-frontier-tier model sibling. In our lag-default sensitivity analysis (Table~\ref{tab:lag-default-sensitivity}), we find that the median \texttt{tier\_gap} under this test is consistently +12.63 \eci{} for all tested values of the lag parameter on the \eci{} scale, and +111.89 Elo on the Arena Elo scale. This is because the dyad-eligible subset is dominated by a handful of within-family tier-sibling model pairs (e.g., Claude 3 Opus and Claude 3 Sonnet; GPT-4o and GPT-4o mini), whose release dates and \eci{} gaps are fixed in the Epoch dataset regardless of how evaluation dates are imputed at the paper level. One effect of changing the lag parameter is to change the size of the dyad-eligible subset ($n = 4{,}645$ at L=0 days; $n = 3{,}641$ at L=365 days): as the lag gets larger, the imputed evaluation date is earlier, so more within-family siblings released later become ineligible.

\subsubsection{H4 -- configuration underreporting (descriptive)}
\label{sec:results:h4}

For papers that evaluate models with reasoning capabilities, we asked: does the abstract explicitly state whether reasoning mode was enabled? Under our primary specification of looking up the capability, only 2.2\% do so (23/1,044; 95\% Holm-Bonferroni simultaneous confidence interval [0.013, 0.036]). Our secondary specification, which differs in that it consistently codes null-flag entries and post-freeze releases by their published capability surface, produces the same rate of 2.2\% (27/1,224; raw 95\% CI [0.015, 0.032]). These numbers are an order of magnitude smaller than the $\geq$50\% falsification bucket we set in our pre-registration.

Out of the 18,574 included papers, we were successfully able to extract machine-readable full text for 4,766 (25.7\%). We found that the full text disclosure rate of each of the applicable configuration components ranged from 71.9\% (prompting strategy) to 5.6\% (tool use). In the full text, the compound disclosure rate (i.e. the rate of papers that disclose all configuration components that apply to their primary model and deployment surface) was 1.18\% (36/3,052; Wilson 95\% CI: [0.85, 1.63]). The full-text compound disclosure rate is highest in the coding (2.42\%) and scientific reasoning (2.16\%) domains, followed by medicine (1.28\%). The compound disclosure rate is 0\% in the education (0/705; 95\% CI: [0.00, 0.54]) and law (0/84; CI: [0.00, 4.37], though note the small n) domains. In terms of year, the compound disclosure rate in the full text is 0.60\% for 2023, 0.95\% for 2024, 1.20\% for 2025, and 3.00\% for 2026. In the full text, 21.2\% (111/524; 95\% CI: [17.9, 24.9]) of papers that evaluate reasoning-capable models disclose the reasoning mode. 5.6\% of papers that evaluate tool-capable models disclose tool use in the full text. In calculating the compound disclosure rate, we only considered papers in which we were successfully able to map the primary model to our per-(model surface) override map, excluding 1,710 records. These results are secondary and descriptive in nature, with the primary (pre-registered) results for H4 and H5 being the headline magnitudes.

The large majority of methods sections do not give an evaluation date. Only 877/4,757 (18.4\%) of successfully extracted results have an evaluation date. The rest are undisclosed and imputed according to our pre-registered policy (\S\ref{sec:methods:frontier}). Whether a paper gives an evaluation date varies substantially by domain and primary model family. For example, 28\% of papers in medicine ($n = 2{,}381$) give an evaluation date, but only 5\% of papers in coding do ($n = 647$; logistic regression odds ratio (OR) = 6.1 comparing medicine to coding, p < 0.001). Papers primarily evaluating models from OpenAI, Anthropic, and Google had 3 to 5 times higher odds of giving evaluation dates than the ``other'' family (OR = 2.98, 4.51, and 4.48, p < 0.001). There is no evidence for a year-on-year trend in eval date disclosure (OR=0.95 per year, p=0.34, HC1 robust SE).

\subsubsection{H5 -- compound failure (descriptive)}
\label{sec:results:h5}

\begin{figure}[!htb]
  \centering
  \includegraphics[width=0.92\textwidth]{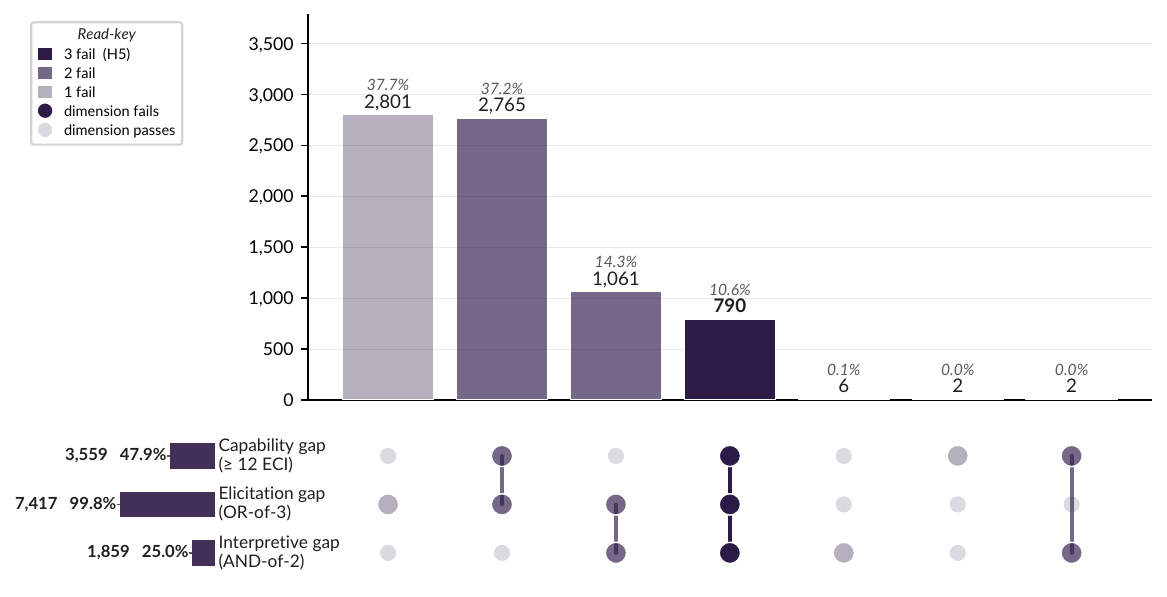}
  \caption{UpSet plot showing the admissibility-expected compound failure rate across medicine, law, education, and scientific reasoning. $n = 7{,}435$ (strict drop-None bars). Left marginals show independent failure rates for each dimension. The darkest bar shows the intersection of all three pre-registered dimensions (790, 10.6\%). Capability gap: $\geq 12$ \eci{} difference using canonical 180-day eval-date imputation. Elicitation gap: OR of three sub-gaps (reasoning, tools, scaffolding). Interpretive gap: AND of two sub-gaps (comparator absence, ai\_generic framing). H5 = 8.9\% (790/8,921; simultaneous 95\% Wilson CI: [8.2\%, 9.6\%]). The pre-registered denominator for H5 is 8,921 papers (those with decidable three-valued AND scores). The strict drop-None bars shown in the figure exclude 1,486 papers from this denominator that had passing and undecidable scores in different dimensions. As a sensitivity analysis, we also computed the compound failure rate when undecidable scores are treated as passing: 790/9,578 (8.25\%) papers failed all three dimensions simultaneously.}
  \label{fig:upset}
\end{figure}

For H5, we provide a conservatively biased descriptive estimate of the frequency of compound failure (Figure~\ref{fig:upset}), which does not have the same epistemic status as the directional sign tests for H1 and H3: we merely report the proportion of papers that compound-fail by various operationalisations (where, conservatively, capability is operationalised by using the mean major generation jump in \eci{} scores as the cutoff, interpretive failure requires both no comparator and a \texttt{conclusion\_framing} of \texttt{ai\_generic}, where either one alone does not suffice; elicitation failure is the inclusive OR of the three disclosure failures). Under the primary operationalisation of interpretive concerns (AND-of-two), 8.9\% of papers where admissibility was expected (790/8,921; simultaneous 95\% Wilson CI [8.2\%, 9.6\%]) exhibited compound failure across all three audit dimensions. Under the alternative, ``inclusive-alternative'' operationalisation (OR-of-two), the frequency of compound failure was bounded above by 36.8\% of papers where admissibility was expected (2,841/7,717). Across the entire corpus, the fraction is 790/17,915 (4.4\%) and 3,677/14,814 (24.8\%), respectively. Table~\ref{tab:h5-spec-grid} visualises all operationalisations and denominators in one grid. Our pre-registration also included a sweep of the capability failure threshold over \{8, 10, 12, 15, 20\} \eci{}, with 12 \eci{} as the primary threshold. The results of this analysis can be found in Appendix~\ref{app:spec}. The rate of compound failure changes smoothly over the sweep and is lower for any stricter threshold. Admissibility is not among the primary fields coded subjectively. In our analysis, it is determined by a categorical rule based on the domain field we extract. To the extent that this rule is noisy, the resulting compound failure rate may be biased in either direction, since a noisier definition of admissibility changes the pool of admissibility-eligible papers and therefore the compound failure rate. The 8.9\% descriptive rate (regarding compound failure in admissibility-expected papers only) is not a confirmatory claim that an effect has been detected.

\begin{table}[H]
\centering
\small
\caption{H5 compound-failure rate across operationalisations and denominators. ``Strict drop-None binary AND'' retains only papers with decidable answers for all dimensions ($n = 7{,}435$ papers). The pre-registered ``tri-valued AND-of-two'' model also allows for papers for which the AND short-circuits to a decidable False. This model has a larger denominator of $n = 8{,}921$ admissibility-expected papers ($n = 17{,}915$ for the full corpus). ``OR-of-two'' uses the inclusive-alternative interpretive arm instead. Primary headline cell in bold.}
\label{tab:h5-spec-grid}
\begin{tabular}{lcc}
\toprule
Operationalisation & Admissibility-expected & Full corpus \\
\midrule
Strict drop-None binary AND        & $10.6\%\ (790 / 7{,}435)$  & --- \\
Tri-valued AND-of-2 (primary)      & $\mathbf{8.9\%\ (790 / 8{,}921)}$ & $4.4\%\ (790 / 17{,}915)$ \\
OR-of-2 (inclusive sensitivity)    & $36.8\%\ (2{,}841 / 7{,}717)$ & $24.8\%\ (3{,}677 / 14{,}814)$ \\
\bottomrule
\end{tabular}
\end{table}

\subsubsection{H6 -- valence asymmetry}
\label{sec:results:h6}

We ran the H6 model (\texttt{eci\_gap} $\sim$ \texttt{conclusion\_valence} + domain + year + domain:year + (1|journal)) on $n = 12{,}660$ papers. For H6 to be considered successful, we need (i) the regression coefficient to be above zero and its confidence interval to exclude zero, and (ii) at least 90\% of the $n = 1{,}000$ simulations from the measurement error model to have an effect with the same sign. The coefficient for the difference between negative and positive valence papers was $\beta$ = $-0.29$ \eci{} (two-sided 95\% CI [-0.79, +0.20], p = 0.25). Because this interval includes zero, the H6 coefficient did not pass (i) and we could not reject the null for the pooled H6 test. We could not informatively check (ii) because the simulation procedure re-estimates the models on the simulated draws using OLS with HC3 robust standard errors and no journal intercept. The same procedure on the observed labels yields a coefficient of $-0.77$ \eci{} (95\% CI $[-1.27, -0.28]$), with a positive sign in 0 of 1,000 simulations. For the interpretive dimension, our main analyses of framing operationalisation (H5) and of the share of claims at the class level (Section~\ref{sec:results:framing}) are both based on \texttt{conclusion\_framing} and do not rely on valence.

\subsubsection{Class-level claim share (descriptive)}
\label{sec:results:framing}

We now turn to the class-level claim share, our chosen descriptive anchor for the interpretive dimension. This is the fraction of conclusions in our sample that make broad claims about ``AI'' as a class, i.e. for which \texttt{conclusion\_framing} = \texttt{ai\_generic}. Per our pre-registered Bayes-corrected estimator (Section~\ref{sec:methods:validation}), the class-level claim share is 52.5\% (simultaneous 95\% confidence interval of bootstrapped marginal posteriors: [47.3, 57.9]). The gold-anchored direct count for the same post-adjudication sample of $n = 231$ papers is 53.2\%.

\begin{figure}[!htb]
  \centering
  \includegraphics[width=0.95\textwidth]{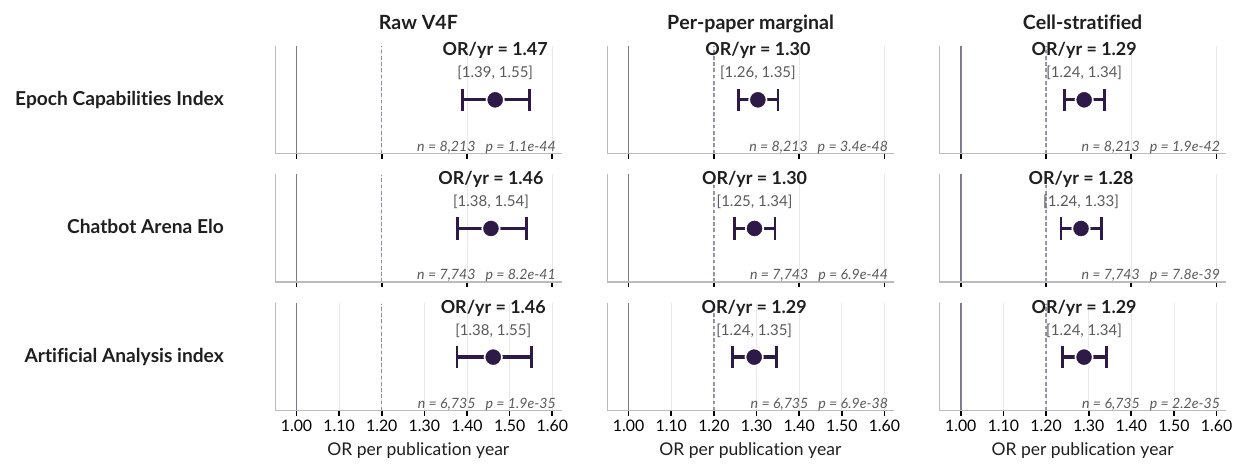}
  \caption{This figure shows per-publication-year odds of class-level (``AI''-framed) abstract conclusions, limiting to the \textbf{five pre-registered domains}. Of the columns, the two Bayes-corrected are primary (raw V4F observation is included just for transparency). Cell $n$ varies from $6{,}735$ to $8{,}213$. Columns denote Bayes-correction specification (raw V4F observation, per-paper marginal posterior, cell-stratified posterior). Rows denote capability coverage subset (Epoch \eci{}, Chatbot Arena Elo, Artificial Analysis intelligence index). Points represent the annual odds ratio and whiskers the nominal $95\%$ confidence interval, from models (logit for raw observations, binomial GLM with HC3 heteroskedasticity-consistent covariance matrix for corrected posteriors) fit on publication year and domain separately in each scale-resolvable subset. Posteriors are computed based on the $n = 231$ post-adjudication-merged dual-coder gold. The solid vertical line shows an odds ratio of 1 (no change) and the dashed an odds ratio of 1.20. Whiskers are not simultaneous confidence intervals.}
  \label{fig:trend-grid}
\end{figure}

The audit's interpretive claim is about the trend over time rather than the level. In the full V4F-cascaded corpus, the estimated odds of framing the conclusion of a paper as applying to the class increase with the publication year. The odds ratio (OR) per year is 1.23 (95\% CI [1.19, 1.27], p < $10^{-33}$, $n = 18{,}565$). We find similar results when restricting to the \eci{}-anchored subset using our primary per-paper marginal posterior specification (OR = 1.24, 95\% CI [1.20, 1.28], p < $10^{-33}$, $n = 12{,}664$) and when restricting to the Artificial Analysis intelligence index subset (OR = 1.24, 95\% CI [1.20, 1.28], $n = 10{,}152$) or the Chatbot Arena Elo subset (OR = 1.24, 95\% CI [1.20, 1.27], $n = 11{,}830$). The grid of specifications shown in Figure~\ref{fig:trend-grid} is limited to our five pre-registered domains. In each case, the Bayes-corrected within-cell odds ratio is around 1.28 to 1.30. These estimates are also above the dashed line showing OR = 1.20.

\subsection{An elicitation-gap exemplar on SWE-Bench-Verified}
\label{sec:results:swebench-exemplar}

We illustrate the effects of compound attenuation in Box~\ref{fig:box1}, and the effects of compound failure at corpus scale in H5 (see \S\ref{sec:results:confirmatory}). In principle, one could instantiate this box for any given task with a series of ablations using matched comparisons that decompose each factor. In practice, such data is rarely reported, and for any given task it is hard to instantiate this waterfall with empirically grounded values. SWE-Bench-Verified is one of the few benchmarks with enough publicly reported results to attempt this. We attempt to ground three out of nine of the axes from Panel B of Box~\ref{fig:box1} (chips 1 to 3) as directly measured on the same benchmark, and bound the remaining six chips (chips 4 to 9) based on the closest public ablations. This allows us to replace the factors stipulated in the schematic in Box~\ref{fig:box1}, Panel B, with a mixture of directly measured and bounded estimates.

\begin{figure}[!htb]
  \centering
  \includegraphics[width=0.95\textwidth]{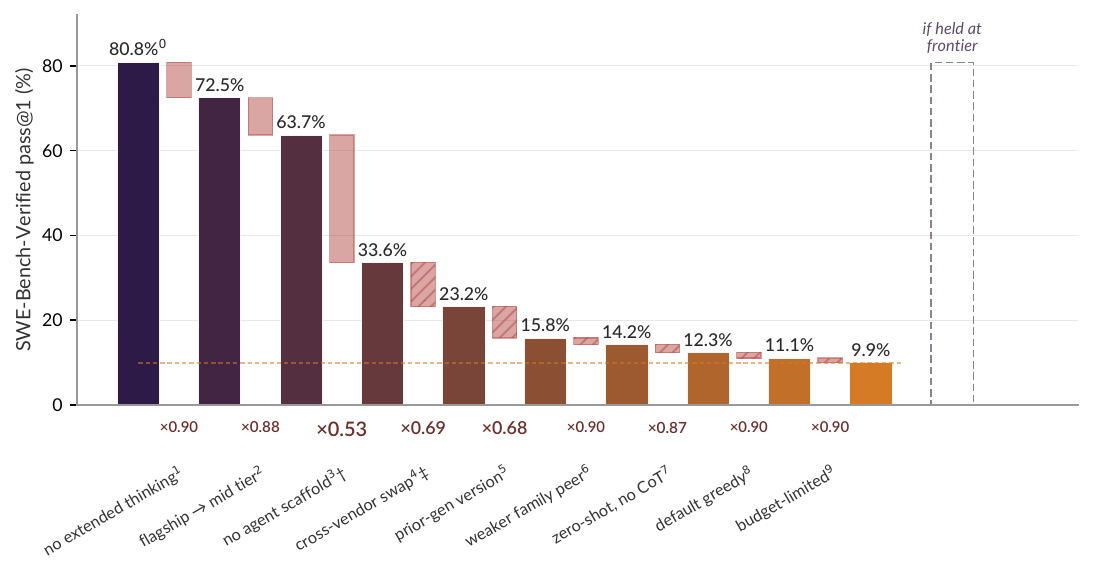}
  \caption{Multiplicative waterfall on SWE-Bench-Verified. Capability falls through nine configuration downgrades, from Claude Opus 4.6 Thinking Max at $C_\text{max} = 80.8\%$ pass@1 to $C_\text{min} = 9.9\%$. (This minimum capability is a stylised low-elicitation value, rather than one we directly measured from our corpus). This implies a compounded retained fraction of capability, after multiplying the effects of each of the nine changes (assuming they are independent axes): $G_\text{total} = \prod_k G_k \approx 0.123$. Chips 1--3 (solid fill) are direct measurements on the same benchmark, also changing model version/generation. Chips 4--9 (hatched fill) are bounded estimates, interpolated from the nearest available public ablation (see Table~S2 for each chip's source). Chip 3 (\textdagger) crosses a generation boundary, from 3.7 Sonnet to 3.5. Chip 4 (\textdaggerdbl)'s ratio is based on a comparison of performance with matched scaffolds between model providers, rather than an ablation of tool use in a single model.}
  \label{fig:waterfall}
\end{figure}

Figure~\ref{fig:waterfall} decomposes a benchmarked public score on SWE-Bench-Verified (Claude Opus 4.6 Thinking Max, with 80.8\% pass@1 (25 trial average) as of 2026-04-23) through nine configuration downgrades to a stylised ``low elicitation'' endpoint (9.9\%), representative of common corpus omissions. This results in a compounded retained fraction (product of per-chip $G_k$) of $G_{\mathrm{total}}$ $\approx$ 0.123 irrespective of chip ordering. Three of the nine chips are based on measured comparisons on the same benchmark; the other six are bounded estimates interpolated from the nearest public ablations. Each of the three measured chips also involves a model version or generation change (chip 1: Opus 4.6 $\rightarrow$ Opus 4; chip 2: Opus 4 $\rightarrow$ 3.7 Sonnet; chip 3: 3.7 Sonnet $\rightarrow$ 3.5). Chip 4 is an interpolated estimate, with the ratio based on a cross-vendor comparison with a matched scaffold rather than a fixed-model ablation removing tools (see Table~S2). This exemplar is for illustration rather than inferential purposes, and is a decomposition of one task in one domain (SWE-Bench-Verified). It is the most complete set of such chips that can be assembled using public information for any benchmark, which is not typical for reported benchmarks. H5 (\S\ref{sec:results:confirmatory}) measures the compound failure rate of the three audit dimensions at corpus scale, which is structurally different from this illustrative exemplar's multiplicative attenuation.

\subsection{Exploratory analyses}
\label{sec:results:exploratory}

Each analysis that follows is exploratory and does not make any $\alpha$-level claims.

\subsubsection{Dispersion structure (H7)}

We applied Hartigan's dip test (H7, registered, descriptive) to \texttt{eci\_gap} in our sample (H1, $n = 12{,}668$), finding a dip statistic of $D = 0.057$ (95\% bootstrap CI [0.054, 0.060] with 10,000 resamples), which was higher than all 1,000 draws from the simulated uniform null ($p = 0.001$, floor). When stratified by year of publication, the $D$ statistic is 0.219 (2023), 0.099 (2024), 0.048 (2025), and 0.072 (2026), with $p$-values at the floor in each case. However, in a post-hoc analysis we found that our measure \texttt{eci\_gap} (a difference of model-level scores) takes 456 unique values, and that two point masses --- corresponding to a gap of 0 (17.2\% of papers, when the evaluated model was the frontier model) and +12.63 (10.5\%, corresponding to GPT-3.5 Turbo vs.\ a frontier GPT-4, with 1,316 of 1,333 papers from 2022-2024) --- by themselves necessitate a $D$ statistic of at least 0.0526, or 92\% of the observed 0.057. Thus, in post-hoc analysis, the dip test result is explained by the discreteness of the measure, and whether there is a continuous second mode in 2025-2026 would require a smoothing specification that was neither registered nor performed. We also find a heavy right tail (by a rule decided in advance: skewness $> 0$ and $(p_{99} - p_{50})/(p_{50} - p_{1}) > 1$) in the pooled sample; within journal (8,213 papers in 721 journals with two or more papers); and in each of 2024, 2025, and 2026. The heavy tail does not appear in 2023, where 95\% of the 1,336 papers have \texttt{eci\_gap} equal to 0 or +12.63, the largest secondary point mass (corresponding to the 2022-2024 GPT-3.5 Turbo atom). The 983 papers from 2025-2026 whose primary model is GPT-3.5 lineage or older have a median \texttt{eci\_gap} of +29.35 (IQR +23.09 to +33.62), compared to a pooled median of +10.45; 614 (62.5\%) are in the pooled top decile (\texttt{eci\_gap} $\geq$ +23.37), where they comprise 48.5\% of the 1,267 papers. This heavy tail, not the dip test result, is the evidence of a genuine feature of a subset of the corpus, rather than an artefact of the corpus construction methodology.

\subsubsection{Publication-lag comparison (H8)}

We cannot simply subtract an external publication-delay benchmark from the median gap of +10.45 \eci{} to calculate how much of it is due to peer review.

Table~\ref{tab:lag-default-sensitivity} shows the robustness to evaluation dates. Where there is an explicit evaluation date in the text, we use that instead of the defaults. In the default 0-day scenario, the median H1 is +14.84 \eci{} (12,671 papers); in the default 180-day scenario, the median H1 is +10.45 \eci{} (12,668 papers); in the default 365-day scenario, the median H1 is +3.83 \eci{} (12,649 papers). These comparisons are based on different date assumptions and also partially different sets of papers that can be analysed, so they should not be taken as a within-paper decomposition or an effect of peer review. It is plausible that model choice also reflects cost and access constraints, but we do not attempt to quantify their impact.

\subsubsection{Measurement invariance across domains (H9)}

In each of the 5 pre-registered domains, as well as the residual ``other'' domain, the directional sign of the effects remains unchanged. The H1 test shows the lowest median for scientific reasoning (+3.09 \eci{}) and highest for education (+12.63 \eci{}); the largest one-sided H1 p-value across domains is $3.12\times10^{-22}$. The H3 test in law shows $p=2.65\times10^{-10}$ ($n = 49$). Note that we conduct these analyses separately by domain, and we do not apply corrections for multiple comparisons. Figure~S2 in Appendix~F shows that H2 slopes are positive within each pre-registered domain and the residual ``other'' domain, which suggests that the gaps are widening over time. In each domain, the H3 median is the modal +12.63 \eci{}, with the exception of scientific reasoning, where it is +9.53 \eci{}. On the primary scale, these are magnitudes for eligible comparisons with a positive gap. The domain-level H3 results should not be taken as independent proof of the presence of this effect. The pooled H6 result is reported in \S\ref{sec:results:h6}.

\subsection{Sensitivity analyses}
\label{sec:results:sensitivity}

\paragraph{Independent-frontier substitutions (Arena Elo, Artificial Analysis).}

H1 medians and H2 pooled point estimates remain positive under Chatbot Arena Elo and Artificial Analysis substitutions for Epoch \eci{}. The class-level claim share trend also remains positive in subsets with coverage on each scale (\S\ref{sec:results:framing}); these regressions use coverage to select papers rather than capability gaps as covariates. The conditional H3 median is positive under the Arena Elo frontier definition (+111.89 Elo) and zero under the Artificial Analysis frontier definition (Figure~S1, Appendix~F). The analysis does not distinguish ties due to score resolution from shared model mappings or other scale differences. Coverage under these alternative LLM capability scores is 63.7\% (11,836/18,574) for Chatbot Arena Elo (which only includes models that joined the Arena at some point) and 55\% (10,192/18,574) for Artificial Analysis.

\paragraph{Cross-family extraction sensitivity ($n = 150$).}

We calculate per-field pairwise Cohen's $\kappa$s across the 3-family cross-extraction panel (n = 150). We also calculate the same value for the dual-human extraction panel to verify that it clears the pre-registered integrity gate based on dual-human $\kappa$ in subjective fields. The dual-human values are listed in Appendix~\ref{app:validation:gold}. The V4F-replacement triad's per-field pairwise $\kappa$s are in Appendix~\ref{app:validation:crossfamily}.

\paragraph{Reported sensitivity analyses.}

The reported sensitivities are not a completed specification curve (we did not test the entire Cartesian product of these choices) nor a permutation test of the specification curve summary statistic. The saved domain and lag-default sensitivities are presented in Appendix~\ref{app:spec}.

\section{Discussion}
\label{sec:discussion}

\subsection{What we found}
\label{sec:discussion:synthesis}

We find that the median paper in our audit trails the frontier by +10.45 \eci{} (H1), or about one ``generation'' of frontier models. This gap widens by +4.07 \eci{} with each additional year of publication (H2, Section~\ref{sec:results:h2}). Only 2.2\% of the 1,044 papers in our dataset that evaluate a reasoning-capable model explicitly state in their abstracts whether reasoning is on or off during evaluation (H4). This is an order of magnitude lower than our pre-registered falsification floor for this hypothesis. We find that 52.5\% of abstracts in our dataset generalise their findings to ``AI'' at the class level, as estimated by our per-paper Bayes-corrected estimator. This is a primary descriptive result of this paper. The odds of making a class-level claim are rising with time (uncorrected per-publication-year OR = 1.23, Section~\ref{sec:results:framing}).

We find that the subset of papers we would expect to be admissible has a compound failure rate of 8.9\% (Section~\ref{sec:results:h5}). H1 and H2 retain their sign on both substitute capability scales and in every imputation cell (Section~\ref{sec:results:sensitivity} and Table~\ref{tab:lag-default-sensitivity}). The median eligible paper trails the higher-tier release within its own family by +12.63 \eci{} (H3). Because our eligibility criterion fixes the sign of the lag on the primary scale, only the magnitude of the lag is an empirical finding (Section~\ref{sec:results:h3}). The remaining tests are largely consistent with our findings, with the exception of H6, our test for valence asymmetry, which does not support our conclusions. We are unable to reject the null hypothesis under V4F, and our pooled estimate from a mixed-effects model is not statistically distinguishable from zero (Section~\ref{sec:results:h6}).

Internal validity is not the focus of our audit: the tests (H1--H6) do not speak to the internal validity of the headline results of a paper, or whether they would replicate if the authors had used a contemporaneous frontier model and all elicitation surfaces. Rather, the audit asks how widespread poor locatability is in our pre-registered corpus of papers from the literature. We found that the models described in typical abstracts are almost a full generation behind the state of the art, that methods sections do not provide much information about the parameters used to tune model behaviour, and that an increasing proportion of papers are making class-level claims. Whether it be a doctor reading an abstract or a policymaker reading a policy brief, they might not look beyond the word ``AI.'' But each audited paper evaluates at least one specific model in at least one specific access tier and elicitation condition. Because our research design compares papers from different publication cohorts, it is unable to identify the effect of peer review.

It is an open question why authors frame abstracts at class level. It could be for epistemological reasons (e.g. believing that the particular model name is representative of a broader class of models), stylistic reasons (e.g. conforming to an abstract template that promotes class-level statements), or strategic reasons (e.g. intentionally vague in order to garner attention). We leave this question aside because, as far as our audit is concerned, the relevant construct is the overall impression conveyed to the reader, and the intended meaning of the authors remains unknown to the clinician or policy reader. Furthermore, the capabilities landscape is diversifying (e.g. with separate model tiers, or separate reasoning modes), which would require more specific statements about model capabilities in abstracts.

Our audit tracks only the \eci{}-gap and extent of disclosure about configuration items. It does not provide a counterfactual assessment of capability. For any given paper it cannot be used to say whether running the experiment on a contemporaneous frontier model, with the full capability elicitation surface, would have resulted in a reversed conclusion. What it does show is that, in the aggregate, the academic literature gives an impression of what ``AI'' is capable of that is increasingly out of date.

\subsection{Implications for downstream consumers}
\label{sec:discussion:downstream}

Beyond the academic AI evaluation literature itself, many other claims are made by referencing this work.

An education technology purchaser might hear that ``AI shows promise in tutoring.'' A clinician might read in the abstract of a procurement report that ``LLMs fail at ECG interpretation.'' An AI policy staffer might read that ``LLMs struggle with legal reasoning.'' In each case, a class-level claim is made, when in fact the underlying evaluation was of a specific model, via a specific level of API access, elicited in a specific way. In the original evaluation study, these class-level claims were made in the abstract, and on a per-paper, Bayes-corrected basis, 52.5\% of such abstracts make class-level claims.

While the methods section may contain sufficient information for a careful reader to ascertain the model subclass used, even here, there is little information provided about the exact parameters of the method tested. Just 21.2\% of reasoning-capable papers disclose reasoning mode; 18.4\% of full-text papers disclose the evaluation date.\footnote{Because of these missing parameters, we built a Python package, \fl{}, that works on a DOI-by-DOI basis. DOIs not in the audit corpus are resolved in real time using CrossRef and OpenAlex. The output is a three-component vector (temporal, tier, configuration), a breakdown of compound failures, and the framing bucket.}

The vector of the three components we identify is a useful analytical object for reasoning about the capabilities reported in any given paper.

For downstream work that relies on claims made about model capabilities in the academic literature, best practice is to condition such claims on the model tier and model configuration used. A flagship model with reasoning turned off is not the same as a scaffolded flagship model. When a citation must be robust to scrutiny with these caveats about model tier and model configuration, the most important thing is to prioritise the subset of the academic literature that more closely aligns with \versio{}.

\subsection{Positive exemplars}
\label{sec:discussion:positives}

See Appendix~H for a list of the six papers we identified that meet \versio{} v1.2 in a bounded-scope manner and a table matching them to the corresponding checklist axis. Our criticism is a structural one; we do not name any negative exemplars, per an asymmetric naming convention in our pre-registration (see protocol \S11). There may be cost or access considerations that influence model selection, but we do not seek to evaluate what causal part they may play in corpus-wide trends; our commitment to not naming negative exemplars is not contingent on any mechanistic explanation of this kind.

\subsection{Implications for editorial policy, funders, and the AI-safety ecosystem}
\label{sec:discussion:implications}

\versio{} enumerates information that should be disclosed to enable readers to reconstruct the configuration of a tested model, and understand the scope of the claims of a publication. The Core 3 disclosure in most papers would require about 500 characters in the methods section (about two sentences), covering the exact model version evaluated (Item 1), the capability frame being claimed (Item 5, congruent with the tier of model evaluated), and whether or not reasoning was used in the evaluation (Item 7, if applicable to the model). Other \versio{} items apply where relevant. This would enable busy clinicians, meta-analysts seeking to synthesise the literature, policy-makers preparing briefing papers, or AI safety analysts tracking capability trajectories, to contextualise the capability claims of a given abstract within the capability trajectory. Potential avenues for operationalising the \versio{} checklist include: requiring Core 3 disclosures in journal submission portals; developing guidance prompting peer reviewers to confirm that claimed configurations are sufficient to support the scope of conclusions; and recommending to funding bodies that they require explicit funding for evaluation and model configuration disclosure in funded publications. More specific avenues could include a CONSORT-AI extension based on elicitation, a TRIPOD-LLM addendum, or a DECIDE-AI addition; the suitability and cost of implementation of these should be assessed with journals and the relevant reporting guideline working groups. Funder support for evaluation access is proposed as an intervention for future work to assess; we do not seek to estimate the likely effects of disclosure, evaluation funding, or shorter publication cycle requirements on the capability gap distribution.

\begin{figure}[!htb]
  \centering
  \includegraphics[width=0.85\textwidth]{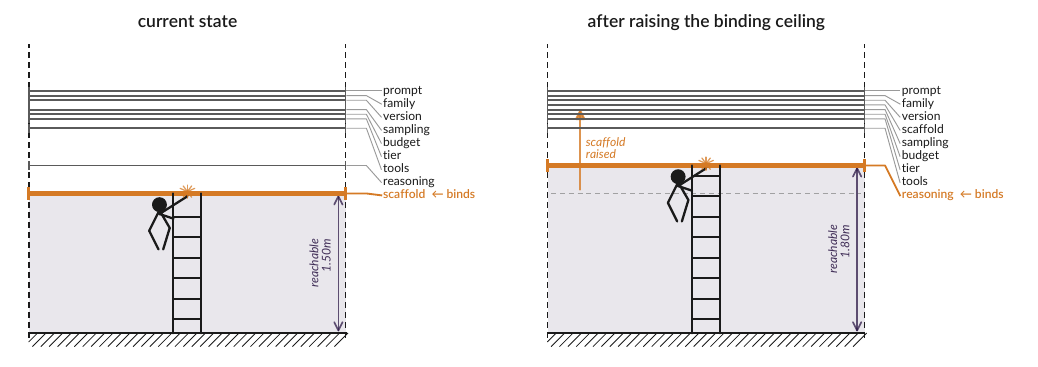}
  \caption{Reachable capability as a ceiling-stack cross-section. Each of the nine configuration axes is associated with a suspended ceiling that limits the outcomes that can be reached. The binding constraint is the lowest ceiling. Left: the binding constraint is scaffolding. Right: with the schematic scaffolding ceiling raised, the binding constraint is reasoning mode.}
  \label{fig:ceiling-stack}
\end{figure}

Per-axis reporting may be seen, within the context of the ceiling-stack shown in Figure~\ref{fig:ceiling-stack}, as raising the ceiling on the most-tractable axis, while the binding constraint remains in the unreported axis. It may be that funder-side interventions will be needed; evaluating frontier capability is expensive. Where a funding grant does not make explicit provision for API access, there is a risk that the academic literature evaluating AI comes to represent a small oligopoly of well-funded labs with close industry ties, able to conduct capability-relevant elicitation evaluations at scale. Grants without API support may also leave independent academic groups evaluating older models accessible with free-tier API access, which are less informative about the capabilities of AI artefacts that policy-makers and clinicians increasingly encounter. Many funders of health research, such as UK Research and Innovation (UKRI), the National Institutes of Health (NIH), and the National Science Foundation (NSF), as well as major private funders, have a vested interest in making sure this divide does not become more acute.

While recent randomised evaluations of AI in medicine were prospectively registered and identified the models being tested \citep{bean2026preregistered, qazi2026lmic}, the gap we identify in our audit occurs downstream, in claims and citations, where the information provided in methods sections meets the inferences drawn by downstream readers.

\fl{} (\url{https://frontierlag.org}) is available as both a web tool and CLI tool. Given any DOI in the audit corpus, it returns the three-component capability-distance vector as defined in this paper, the compound-failure decomposition as defined for the AND-of-two operationalisation, and the \texttt{conclusion\_framing} bucket. It also resolves DOIs outside the audit corpus using CrossRef and OpenAlex. We keep the pre-registered thresholds constant through each quarterly update to the Epoch AI trajectory.

\subsection{Open questions and the VERSIO-AI v2 comment period}
\label{sec:discussion:open}

\versio{} v1.2 will be open to community feedback and input within 60 days of this version. We anticipate that the feedback will lead to changes in item 5 (the ``declared capability frame'') and item 12 (sampling and determinism reporting) in particular. A companion specification details the item-revision protocol to which we are committing. As an important next step, we plan to bring non-English-language literatures into the audit, and we plan to do so in V2 of the \fl{} dataset. If domain-specific capability indices (indices of clinical reasoning, long-horizon coding, and legal citation capabilities) become public and their methodology is open, we plan to incorporate them into \fl{} v2. In 2028Q2, a follow-up audit will re-run the same protocol to assess whether the distribution across the corpus has changed.\footnote{Frontier providers could subsidise academic access; journals could adopt \versio{}-style reporting; pre-registered reviews of AI-evaluation papers could become routine. Whether all of this shrinks the distribution is a question for data, not speculation. A follow-up audit in 2028-Q2 would measure how the distribution has changed; linking those changes to adoption would require a separate causal design.}

\section{Limitations}
\label{sec:limitations}

This section lists, from most to least load-bearing, the problems that our validation protocol does not resolve.

\subsection{Corpus-selection bias}
\label{sec:lim:corpus}

OpenAlex is not a census of research output: non-English research, some conferences, and ``grey literature'' are under-represented. We do not model the effect of these biases on our estimates of the distribution of the \eci{} gap. However, it is possible that these biases are especially prevalent in regions with low frontier access. The coverage audit reported in Section~\ref{sec:methods:corpus} provides our best guess about the extent to which this is the case (though we do not view this as a bound). To the extent that coverage issues are more common in regions with low frontier access, our findings should be treated as within-topic audit results. The within-topic coverage audit reported in Appendix~G finds our title keyword-based approach has identified about 60\% of LLM evaluation papers in two OpenAlex concept topics our audit subsumes. Based on the frontier-gap proxy, the missed sample has a higher median (+10.7 vs.\ +10.3 months) and mean (+12.6 vs.\ +9.9 months) gap to the frontier than the corpus, meaning that missed papers do not seem to be closer to the frontier. The two proxy distributions are statistically significantly different at an uncorrected significance level of 0.05 (Mann-Whitney $p = 0.0043$), but not with a Bonferroni correction for $k = 18$ tests. Other than the overall inclusion rate, only the compositional cells survive the correction. Of these, we see that the missed sample is overrepresented on the \texttt{chatgpt} and underrepresented on the \texttt{gpt-4} abstract primary-model tokens (as opposed to title keyword matches). This is consistent with the under-specification structure we identify and document in our in-corpus sample.

Full-paper extraction is limited by PDF accessibility within the OpenAlex database, which is constrained by paywalls and licensing. Full-paper extraction could be attempted for 4,766 (25.7\%) out of 18,574 papers. Because we only had access to abstracts for the larger set of papers, we performed the audit at the abstract level. The set of papers with a PDF accessible is a non-random subset. Broken down by classifier-assigned domain, they are more likely to belong to the fields of medicine (50.4\% vs. 28.4\%) and education (17.3\% vs. 7.8\%) and less likely to belong to the ``other'' catch-all domain (4.0\% vs. 49.8\%). They are also more likely to be from 2023 and 2024, and less likely to be from 2026. However, they make abstract-level class-level claims at a similar rate (40.6\% vs. 42.8\%), meaning our load-bearing interpretive-dimension finding is robust to this sample split.\footnote{At the abstract level, the papers in the retrievable and non-retrievable subsets disclose the evaluation date (3.0\% vs. 2.6\%) and configuration fields (prompting strategy 21.0\% vs. 21.8\%) at rates less than one percentage point apart.}

Our human validation focused only on the precision of the classifier, because the $n = 450$ gold-standard sample was drawn from the pool that the classifier predicted as includes. Relative to the boundary of AI evaluation, the classifier's recall is only estimated, but this could be bounded more precisely using a recall-aware human validation sample that also sampled papers from the predicted-exclude pool of the classifier. However, sampling on the exclude side requires its own pipeline, so this would be a separate methodology project from the audit pipeline validation reported in this paper. We have limited this recall-aware validation to version 2 of the audit, and in version 1 we rely for evidence of coverage on the Bonferroni-corrected comparison between the corpus and the residual pool (Appendix~G), which does not check for recall on a paper-by-paper basis, and which we report as a within-topic coverage bound rather than a global bound. Headline magnitudes are based on $n = 12{,}668$ papers.

\subsection{Extraction-pipeline concentration}
\label{sec:lim:pipeline}

We use a single model, DeepSeek V4F-Max, to both classify papers for inclusion and extract all subjective fields. This refers to DeepSeek V4-Flash-Max, used at its maximum reasoning setting. For the rest of the paper we refer to this as V4F. Two human coders labelled the $n = 231$ gold standard which we used to validate V4F. We report, field-by-field, the extent to which V4F agrees with the gold standard in Table~\ref{tab:benchmark-four-extractor}. We also conduct another run with our frozen prompt on a stratified subsample of papers ($n = 150$, see Appendix~\ref{app:validation:crossfamily}). This run uses two triads of models, where models within a triad are selected from different model families. The first triad (a) is our pre-registered triad: gpt-5.4-mini, gemini-3.1-pro-preview, and claude-opus-4-7. The second triad (b) replaces gpt-5.4-mini with DeepSeek V4F-Max; thus the models in this triad are gemini-3.1-pro-preview, V4F, and claude-opus-4-7.

All load-bearing subjective fields have $\kappa$ $\geq$ 0.65 between V4F and Opus in triad (b). We note that agreement between LLM-as-measuring-instruments cannot rule out failure modes shared by transformer-family LLMs. This is a downside of our design: our object of study and our measuring instrument share some failure modes in common. On the other hand, it is part of the thematic point of this pipeline that a frontier LLM is measuring the scientific literature on frontier LLMs (Section~\ref{sec:methods:extraction}). A human panel coding all 112,303 papers, matched to the model tier of the LLMs, would be well outside the budget of this paper. Where it is important, such as for \texttt{conclusion\_framing} (Section~\ref{sec:results:framing}), we take the analytic-correction route we pre-registered.

\subsection{Bayes-correction transportability}
\label{sec:lim:bayes}

The framing-field correction (Section~\ref{sec:methods:validation}) uses a small anchor ($n = 231$ dual-coded gold pairs) and residual errors matter here: the relevant error envelope for the headline 8.9\% rate is given by V4F's per-field $\kappa$ against the gold standard (Table~\ref{tab:benchmark-four-extractor}). One might conduct a fully Bayesian treatment, sampling from uncertainty about each field's $\kappa$ and propagating that uncertainty to the AND-of-two conjunction on which H5 relies. We leave this to future work; this is unnecessary for the argument for v1, which instead hinges on Section~\ref{sec:methods:outcomes}.

Our pre-registered framing-extraction prompt is not perfectly stable. On our development set, two temperature-0 re-runs gave the same framing 88.7\% of the time, though this noise is not captured by our bootstrap procedure. Regardless, the trend over publication years holds.

Only 2 gold pairs fall into the catch-all ``other'' category, so we report an uncorrected rate for that domain. All 18,565 papers receive this correction in the pooled 52.5\%, or 48.5\% if we left ``other'' uncorrected. Regardless of the estimator used, the corpus rate is within an 11 percentage-point window. The confidence intervals for the individual domains overlap, given a gold n of 231. In all cells, the trend holds (Figure~\ref{fig:trend-grid}).

\subsection{Elicitation-axis audit is partial}
\label{sec:lim:elicitation}

Six of the eight items (reasoning mode, thinking effort, tool use, scaffolding, multi-agent architecture, and prompting strategy) were operationalised as binary disclosure flags, with the remainder (access method and temperature) operationalised as descriptive readouts (Methods~\S\ref{sec:methods:extraction}). There may be other important ways in which researchers' evaluations of frontier model capabilities are systematically less than what models are truly capable of. For example, the quality of prompting strategy can make a difference in evaluated model capabilities. This includes the choice of in-context exemplars, the design of chain-of-thought templates, and the use of role prompting. These disclosure flags are only able to capture whether or not the authors describe their prompting strategy in their paper, and cannot measure the quality of the prompting strategy itself. Similarly, language model decoding parameters like temperature, top-p, max-tokens, and seeds can cause changes in performance large enough to change the rankings of state-of-the-art models \citep{hochlehnert2025sober}. Finally, choice of judge model for LLM-as-judge evaluations is another capability-relevant factor, which is not tracked in this audit. These factors can all worsen evaluations without tripping a disclosure flag, meaning that the elicitation-dimension rate reported here misses part of the true elicitation deficit. Researchers who pass these H4 disclosure flags could still be substantially worsening the evaluated capabilities of models by using lower-quality prompting strategies, different decoding parameters, or different judge models.\footnote{The audit measures reported elicitation conditions rather than latent capability under controlled elicitation, the latter addressed by a separate literature on evaluation-format and scaffold sensitivity \citep{sclar2024formatspread, pezeshkpour2024sensitivity, gringras2026sus}.}

\subsection{Multi-model papers reduce to a single primary model}
\label{sec:lim:extraction}

This metric extraction scheme includes a field for \texttt{primary\_model}. However, some papers in the corpus evaluated multiple frontier models. We use the highest-\eci{} model evaluated in each paper as the primary model for paper-level aggregate analyses. We also create a per-model dyad file that captures evaluations of every \eci{}-scored model, as well as a multi-model sensitivity analysis. Our paper-level analyses may overstate the capability surface of papers whose primary claims rest on the evaluation of a non-primary model (as defined above).

\subsection{Valence coding is subjective and H6 access-covariate is omitted}
\label{sec:lim:valence}

Valence (in the H6 mixed-effects regression) is a 4-category LLM-based label. Two human coders agreed with each other at $\kappa$ = 0.767; the label's agreement with their adjudicated gold is lower ($\kappa$ = 0.723, $n = 229$ of 231), below the 0.75 subjective-field floor. If these are systematically miscoded in a way that is correlated with model age, then the H6 estimate may be biased in either direction. Besides coding accuracy, the H6 mixed-effects regression omits the pre-registered \texttt{author\_affiliation\_type} (our proxy for author access; see our deviation register \S\ref{sec:lim:deviations}). We do not rely on H6 for our interpretive-dimension claims, instead relying on the share of class-level claims (\S\ref{sec:results:framing}) and the H5 compound-failure rate, which focus on \texttt{conclusion\_framing} rather than valence.

\subsection{H1 effect-presence floor and conditional H3 magnitude}
\label{sec:lim:h1-floor}

The H1 claim about magnitude is the core claim for H1. Under our pre-registered 180-day imputation default, we find the median H1 lag is +10.45 \eci{}. Across lag-default scenarios, this varies from +14.84 to +3.83 (Table~\ref{tab:lag-default-sensitivity}). The H1 median varies from +10.45 (imputed-anchor headline) to +3.52 (disclosed-at-eval-date subcorpus; \S\ref{sec:results:confirmatory}). The H1 claim about distributional form is another core claim for H1. The claim of rejecting the H1 null is near-tautologically true. There are a large number of papers (n$\approx$12,668) for which we can compute an \texttt{eci\_gap}, and our pre-registered H1 null is a structural zero. Under these conditions, any non-trivial positive-location result will near-tautologically reject the null hypothesis. H3 further has the limitation that, due to requiring a higher-\eci{} sibling, all primary-scale \eci{} gaps must be positive by construction, and the H3 test can only be interpreted as an analysis of the median and distribution of these positive primary-scale \eci{} gaps conditional on this selection effect. This argument about sign does not apply to our Arena and Artificial Analysis differences, which do not have sign guaranteed by an \eci{} eligibility predicate. The reason we pre-register structural zero null hypotheses is to pre-commit to specific tests before looking at the data, in order to avoid the risk of tuning our hypothesis test thresholds post-hoc. H1 has a near-tautological presence-of-effect floor in the manuscript. In the manuscript, we retain the pre-registered H3 computation and the original Holm family, but interpret the primary-scale H3 as a conditional descriptive magnitude rather than a non-tautological presence-of-effect test. We instead rely on the magnitudes and sensitivity analyses (Appendix~\ref{app:spec}) for the substance of our claim.\footnote{Readers interested in effect size should read the magnitudes; the Holm-adjusted p-values carry the presence-of-effect claim only, and the primary-scale H3 p-value carries no independent presence-of-effect claim.}

\subsection{ECI as single-index scalar}
\label{sec:lim:eci}

Any single-index measure of the frontier inevitably flattens a multi-dimensional profile of capabilities into a scalar. Epoch is transparent about some of the resulting limitations of their index, noting for example that more narrow models ``may receive low \eci{} scores, despite being very capable within their domain'' \citep{epochai_eci_methodology}. Of the 12,668 H1 primaries, 11 are coding-specialist models and none is a clinical, legal, or educational fine-tune. The capabilities index only allows for relative comparisons between models rather than standalone claims of absolute capability (see Section~\ref{sec:methods:frontier}, Appendix~\ref{app:eci}). Scale-dependence of the findings matters; the audit includes cross-scale tests of our hypotheses, but does not endorse a ``winner.'' We replicated H1 and H2 signs using the Chatbot Arena Elo scale, an independent definition of the AI capabilities frontier based on human preference in head-to-head matchups, rather than benchmark aggregation. We also replicated these findings using the Artificial Analysis intelligence index, another independent definition of the AI capabilities frontier based on benchmark aggregation. We replicated H3 using the Chatbot Arena Elo scale, but not the Artificial Analysis intelligence index. We encourage readers to use the three-component vector of time, tier, and configuration information to make their own decisions about how to weight these components, compared to our pooled summary. By pre-registration, we did not require any confirmatory sign to pass the test of all alternative scales. We acknowledge but do not correct for the issues of training-time benchmark awareness and frontier-scale information compression in benchmark-aggregation-based AI capabilities indices. We justify our decision to use this general-purpose anchor rather than a domain-specific one in Appendix~\ref{app:eci}; this is a key load-bearing decision in our audit.

\subsection{Class-framing severity is binary}
\label{sec:lim:framing-severity}

A generic capability claim can be supported by a broad panel of frontier models, or by a single weak-tier model under sparse capability elicitation. These claims would have very different evidential strength. In the binary coding of our audit (\texttt{ai\_generic} vs. \texttt{model\_specific}), we abstract away from this distinction to preserve the structure of our pre-registered framing map. In a v2 framing audit, severity tiers could be indexed by the number of models, diversity of model families, proximity to the capability frontier at the time of evaluation, and the capability frame claimed by the paper. In this v1 audit, the 52.5\% prevalence of generic capability claims is an unweighted prevalence over these unknown severity strata.

\subsection{Bibliometric construct versus individual-author claim}
\label{sec:lim:bibliometric}

The primary output of our audit is a bibliometric measure of the distance between a capability claim and the frozen Epoch benchmark trajectory, on both a paper-by-paper and corpus-wide basis. Though well-defined relative to \eci{} and the frontier benchmark trajectory, it is not a measure of misrepresentation on the part of individual authors. We did not design it to be such a measure: our positive exemplars were intentionally asymmetric (\S\ref{sec:discussion:positives}). We do not make any claims about the good faith or personal judgements of authors. We leave unresolved the exact causal contribution of our proposed structural factors (reporting practices, model access, timing of publication).

Readers interested in the reporting surface of specific papers can refer to the per-DOI audit on \fl{}. Whether the headline result of a specific paper would reverse, were the paper to be re-evaluated using contemporaneous frontier models, is a question for future replication work.\footnote{While our audit describes structural aspects of reporting that could influence the interpretation of the paper for future readers (\S\ref{sec:discussion:downstream}), it does not measure this interpretation, nor does it trace a single chain of citations. Tracing a single chain of citations would require us to identify the papers in the chain, which falls outside the structural focus of our audit.}

\subsection{Deviations from the pre-registered protocol}
\label{sec:lim:deviations}

Smaller deviations from the pre-registration, which do not meaningfully impact any analyses or interpretation, are recorded in a formal register of timestamped deviations, available from the authors on request. We report the following deviations from the pre-registered analysis plan.

The originally pre-registered data extraction procedure specified the use of gpt-5.4-mini. In the final analyses, gpt-5.4-mini was replaced with V4F-Max for all inclusion classification and subjective field extractions, on cost-coverage grounds (see Methods, \S\ref{sec:methods:extraction}). The validity of this model swap is supported by the 4-extractor benchmark tests and the cross-family extractor triad analysis. In addition, a per-pair $\kappa$ analysis specifically for the gpt-mini cross-family triad, which is anchored to the pre-registered gpt-5.4-mini extraction, is available from the authors on request.

H5's registered task-level admissibility rule (protocol \S4.6) was implemented by domain: only medicine, law, education, and scientific reasoning papers were admissibility-expected. Our model lookups extend the registered release-date table with entries added after registration, including Llama, Qwen, and GPT-4.1 models. The pre-registered analysis plan for hypothesis 6 included the author affiliation type and venue type as covariates in the mixed-effects model. However, these covariates were ultimately not included in the final model. Neither field was present in the production analysis frame. The pooled H6 test does not reject the null.

The pre-registration stated that we would extract data only at the abstract level. After pre-registration, we added an additional data extraction pass using two hardened companion prompts on the full text of papers (where machine-readable PDFs were available; $n = 4{,}766$). We did this for two reasons. The first was to implement our evaluation date imputation policy. We used a deterministic forbidden proxy filter to exclude nine types of non-evaluation dates, specifically dates corresponding to submission, acceptance, publication, copyright, training cutoff, model release, benchmark publication, dataset collection, and prior study. We used this filter to gate model-based extraction of evaluation dates from the paper's full text. The second was to compute disclosure rates and descriptive statistics for the full text of papers in addition to the abstracts. These appear as secondary descriptive statistics (see \S\ref{sec:results:confirmatory}). The binding pre-registered primary descriptive statistics are those for abstracts.

An example of this third type of deviation is the analysis of reasoning mode disclosure. As pre-registered, the primary analysis was restricted to abstracts. That said, in a secondary analysis limited to papers with accessible full text and models with reasoning capabilities, the rate of disclosing reasoning mode was 21.2\%.

We also see an increase in class-level conclusion framing (\texttt{conclusion\_framing}, per-year OR = 1.23; see Section~\ref{sec:results:framing}). This is a within-corpus relationship with respect to publication years. We examined the performance of our extractor on the 222 gold papers also labelled by V4F. The fraction of these papers with the human-applied class-level code of \texttt{ai\_generic} rose from 42.4\% for papers published in 2022--2024 to 59.9\% for papers published in 2025--2026. For \texttt{ai\_generic}, V4F had sensitivity of 0.75 and specificity of 0.96 for earlier papers, and 0.76 and 0.95 for later papers.

The confirmatory gold set of sixty human-included papers per domain was not realised, because human inclusion fell below sixty in four of the five domains; reliability and correction use the 177 both-included and 231 post-adjudication papers instead (\S\ref{sec:methods:validation}). Disagreements were adjudicated by an LLM applying the coding rubric under the direction of D.G. (one of the two coders), rather than an independent third reviewer as in the registered protocol. The adjudication log, registered for deposit, is instead available only upon request. If we use either coder's labels as a reference standard, the corrected class-level share is 50.2\% or 58.6\%, rather than the adjudicated 52.5\%. The task-description taxonomy was merged from twelve to ten classes after the twelve-class agreement ($\kappa$ = 0.692) fell below its 0.75 floor. The pre-registered specification curve, its permutation null, the domain-stratified capability elicitation shortfall, and the H10 deployment-frontier test are not reported; Appendix~\ref{app:spec} reports the lag-default and domain-stratified sensitivities. The registered $\pm$90-day sibling imputation of missing \eci{} scores (protocol \S4.1) was not implemented; on our model tables it would have imputed no score.

We have lower inter-coder reliability on our primary model than planned ($\kappa$ = 0.681; it should be at least $\kappa$ = 0.80), but when we ran the gate, we got $\kappa$ = 0.896, so we missed the stop (see protocol \S7.2). We have updated our coding scheme to be consistent with our registered rule for this gate, but we did not re-validate our coding scheme with two coders as specified in our registered protocol. This was out of sequence (the re-validation should have occurred before continuing with production extraction). See Appendix~\ref{app:validation} and Table~\ref{tab:kappa-7-2} for disclosure and replacement checks. Our registered definition of the deployment-accessible frontier is models with inference cost no greater than 10$\times$ that of the cheapest model in the frontier tier on the evaluation date. In our operationalisation (Section~\ref{sec:methods:frontier}), we instead measure the deployment-accessible frontier as models with inference cost no greater than 10$\times$ that of the cheapest model in the non-frontier or ``base'' tier. We made this change because at any given point, prices within the frontier tier vary by less than an order of magnitude and would trivially be within the threshold. No reported result depends on this, as H10 is the only analysis that uses the deployment-accessible frontier, and we did not end up running H10. In the first version of this paper, we did not run the registered Hartigan dip test under H7. In this version, we test for a dip using diptest (version 0.11.0). We fit the test on our sample from H1 ($n = 12{,}668$). Compared to a null of a uniform distribution (1,000 draws), we find a Hartigan dip statistic of $D = 0.057$ [0.054, 0.060], $p = 0.001$.

We do not enforce the registered objective-field halt from \S7.3, which required the minimum exact agreement between all three predictions to be $\geq$0.80. The extracted primary model has exact agreement of 0.792 (vs. 0.790 in the pre-registered model triad), while the primary model in this analysis (selected by the pre-registered highest-\eci{} criterion) has exact agreement of 0.851 (vs. 0.820 in the pre-registered model triad). We report the halt in Appendix~\ref{app:validation:crossfamily} and relevant bounds under H1 in Appendix~\ref{app:validation:gold}.

\subsection{Temporal generalisation}
\label{sec:lim:time}

The monthly trend as estimated by Epoch was about +12.5 \eci{}/year between 2023-03 and 2026-04, so the widening in H2 (+4.07 \eci{}/year) was about one-third of that overall rate. All gap distributions are with respect to the dataset freeze (2022-01 to 2026-04), and which papers they describe in 2027 will depend on how the unstable race dynamic plays out.

\versio{} v1.2 might spur norms of auditable reporting that become widely shared by editorial boards and journals. Some editors and funding sources operate on longer time scales, but by 2028, they will face the question of whether typical AI evaluation submissions are sufficiently annoying or crucial to scientific progress that a journal-level reporting checklist is needed.

This analysis does not reflect the two further frontier releases since the corpus closed on 2026-04-01: GPT-5.5 on the API on 24 April 2026 \citep{openai2026gpt55}, and DeepSeek V4 Pro, an open weights release the same day \citep{deepseek2026v4pro}. Papers published in the coming year will lag behind these releases.

These new releases have no effect on any of the numbers reported in this article. Already, the median paper in the literature was about 10 \eci{} behind the frontier. Because \fl{} is computed for each DOI relative to the contemporaneous frontier, the \fl{} metric for a given paper will tend to be less out-of-date than the metric from the manuscript. However, we cannot know, based on this data alone, whether the distribution of \fl{} will shift. In 2028-Q2, we will conduct another audit to assess this.

\section{Conclusion}
\label{sec:conclusion}

\subsection{What we showed}
\label{sec:conclusion:what}

Across the audited papers ($n = 12{,}668$), the AI system the median paper reported testing was +10.45 \eci{} less capable than the frontier at the time (Section~\ref{sec:results:h1}), a gap 1.3 times as large as the difference in capability between Claude 3.7 Sonnet and Claude Opus 4.5. This gap is widening by +4.07 \eci{} per year (Section~\ref{sec:results:h2}). Among papers that tested reasoning-capable AI, 2.2\% disclosed whether reasoning mode was used in the abstract and 21.2\% in the full-text (Section~\ref{sec:results:h4}). Among the audited abstracts, 52.5\% had class-level claims, and the odds of observing class-level claims in the abstract increased with an uncorrected per-year odds-ratio of 1.23 (Section~\ref{sec:results:framing}). Other results are reported in Section~\ref{sec:results:h3} (tier lag), Section~\ref{sec:results:h5} (compound failure rate), and Section~\ref{sec:results:sensitivity} (sensitivity analyses).

When a clinician does read the methods section to try and see what model was tested, they often cannot find the information they need to identify the model. It is also often not possible to recover the elicitation surface used. When abstracts do not include information identifying the AI tested, clinical, regulatory, and policy papers that rely only on these abstracts also lack this information. We focus on the population-level trends identified in this audit, and do not mean to criticise the individual decision of any specific paper.

\subsection{What changes if the audit is acted on}
\label{sec:conclusion:ask}

Core 3 compliance is proposed as a requirement for avoiding desk rejection. This would require the addition of only about two sentences to a typical methods section. These sentences would describe the specific model version evaluated (\versio{} v1.2 Item 1), its capability frame consistent with the capability tier under evaluation (Item 5), and where applicable, the reasoning-mode status (Item 7). The other ten \versio{} items would be included where applicable.

We do not model the effects of different combinations of layers of adoption over different timeframes on the literature's distribution of outcomes. Research not reliant on grants would not be affected by funder conditioning (one proposed layer of adoption). However, we do not know precisely what share of research this is, because the V4F pipeline does not extract paper-to-funding links.

\subsection{What we are opening}
\label{sec:conclusion:v2}

Three aspects of this work are open: our pre-registration, our open-source software (\fl{}), and this paper itself. The data bundle is available from the authors on request. \versio{} v1.2 is a proposed reporting specification, and we will begin a 60-day period for community comments upon the pre-print of this paper being deposited on arXiv. The item-revision protocol is provided in the companion spec. We hope that \versio{} can be integrated into existing AI reporting frameworks (such as CONSORT-AI, TRIPOD-LLM, DECIDE-AI, STARD-AI, and extensions to SPIRIT-AI), but it can stand alone if this is not possible. We welcome recreation of this audit, forking of the code-base, domain extension to other languages and literatures, and hostile re-auditing under different capability scales. In the course of such re-auditing, it is not necessary that all confirmatory signs of the present audit be upheld. Indeed, where they are not, they provide useful information for the next version of this specification. We will conduct a follow-up audit using the same protocol in 2028-Q2, based on the same corpus construction rules, to test whether this pattern has shifted.

\clearpage
\appendix
\section{VERSIO-AI v1.2 checklist}
\label{app:versio-ai}

The table below presents the titles of the 13 items in \versio{} v1.2. The separate specification for \versio{}, available at \url{https://doi.org/10.5281/zenodo.20060459}, provides a rationale and worked examples of good and bad reporting for each item. The items have the same numbers as in this appendix, but the specification and its associated journal-portal template have a different desk-reject rule from the Core 3 proposed here: the specification and template use Items 1 and 3 (model version and evaluation date) as the desk-reject triggers, whereas the Core 3 proposed here uses Items 1, 5, and 7. We have not yet reconciled these two versions of the desk-reject rule.

\begin{center}
\footnotesize
\begin{tabular}{@{}r c >{\raggedright\arraybackslash}p{0.68\textwidth} c@{}}
\toprule
Item & Block & Title & Core 3 \\
\midrule
1 & A & Model version, to the exact identifier the provider exposes & $\checkmark$ \\
2 & A & Provider and access method &  \\
3 & A & Access or evaluation date window &  \\
4 & B & Within-family tier and rationale for tier selection &  \\
5 & B & Declared capability frame (frontier / deployment / tier-specific), coherent with the tier identified under Item 1 & $\checkmark$ \\
6 & B & Comparator presence, type, and version &  \\
7 & C & Reasoning mode status, where applicable & $\checkmark$ \\
8 & C & Reasoning effort or thinking budget, where applicable &  \\
9 & C & Tool use and retrieval &  \\
10 & C & Scaffolding, agent framework, and multi-turn structure &  \\
11 & C & Prompting strategy &  \\
12 & D & Sampling parameters and number of runs per item &  \\
13 & D & Conclusion--evidence concordance and valence-conditional caveats &  \\
\bottomrule
\end{tabular}
\end{center}

The Block column identifies which block each item belongs to: A for model identification, B for tier and comparator context, C for configuration and elicitation, and D for evaluation and interpretation. We propose the desk-reject tier to be Core 3. That is, if a paper fails any of Items 1, 5, and 7, then regardless of its performance on other items, it cannot be audited as a capability claim.

There are three dimensions to the audit (capability, interpretive, elicitation), and each has a single Core 3 item. Item 1 requires the model identifier as named by the model provider. These identifiers (e.g., \texttt{gpt-5.4-mini} or \texttt{claude-opus-4-7}) communicate the model family and model tier, so this subsumes Item 4's tier identification. Item 5 requires that the capability frame reported in the paper is coherent with the model tier as identified by the model name. For instance, a paper on \texttt{gpt-5.4-mini} reporting a ``frontier'' capability frame would fail Item 5, but one reporting a ``deployment'' capability frame would pass (even if the abstract does not use the term ``frontier''). Reasoning mode is the sole elicitation item for desk-reject, Item 7, because it is the part of model elicitation that most post-2024 papers omit. This item only applies when the model being evaluated has a reasoning mode, which as of 2026 includes essentially all flagship models (GPT-5 series, Claude 4 and later with ``extended thinking,'' Gemini 2.5 and later with ``thinking,'' Grok 4 and later with ``think,'' and the DeepSeek R-series). The other elicitation items (effort, tools, scaffolding, prompting) are still included at the full checklist level.

The reasoning-era subset is those papers that evaluate reasoning-capable models ($n = 1{,}044$ at the abstract level using the model lookup, which excludes GPT-5 and Gemini 2.5 Pro; $n = 524$ at the full-text level using the per-surface capability map, which includes them). For pre-reasoning papers, there is nothing for Item 7 to apply to, so only Items 1 and 5 are scored.

\section{Construct validity of the Epoch Capabilities Index}
\label{app:eci}

H1, H2, H3, and the capability arm of H5 are directly scored on the \eci{}-gap. H6 models \eci{}-gap as the outcome with \texttt{conclusion\_valence} as the predictor of interest. H4 (denominator: reasoning-capable articles) and class-level claim share (denominator: included articles) are outcomes on the reporting surface and therefore not scored against \eci{}-gap. In \S7.6 of our pre-registration protocol, we commit to documenting the construction and limitations of the index to the extent merited by the importance of the index.

\paragraph{Construction.} The index is anchored to Claude 3.5 Sonnet (130) and GPT-5 (150) \citep{epochai2024,epochai_eci_methodology}, with all other models scored relative to these anchors. The index draws on data from Epoch AI's benchmark grid, spanning five clusters (\texttt{coding}, \texttt{math}, \texttt{agentic}, \texttt{knowledge}, \texttt{writing}). Epoch AI uses item-response theory (IRT) to jointly estimate benchmark difficulties and model abilities. Benchmark scores are rescaled such that random-guess performance is zero.

For each benchmark-model cell, \eci{} takes the maximum score achieved by the model on that benchmark across evaluation settings (e.g., thinking effort levels, inference providers). This per-cell max is what's used for cluster aggregation. The analysis uses a frozen snapshot of the data (April 2026), available on OSF (\texttt{data/eci\_scores.csv}; SHA-256 hash reported on OSF). Also available on OSF (\texttt{data/epoch\_benchmarks.csv}) are the underlying benchmark-level cells of the \eci{} data for building alternative capability indices on the same benchmark grid.

\paragraph{Alternative-weight sensitivity.} We pre-registered Spearman rank correlations between \eci{}-gap and three alternative cluster weightings: equal weight per cluster, just coding and math clusters, and just knowledge and writing clusters. We don't report those here. In the main text, the comparison to Arena Elo (see below) is used as the main external validity check.

\paragraph{External-benchmark correlation.} There are $n = 53$ models in common between the frozen April 2026 Epoch data and the Chatbot Arena. Across these models, \eci{} and Arena Elo correlate at Pearson $r = 0.934$ (95\% confidence interval using Fisher $z$-transform: [0.889, 0.962]); the Spearman rank correlation is $\rho = 0.918$. This is well above the pre-registered threshold of 0.80, which we said we would use as a prompt to discuss \eci{}-Arena decoupling. In Figure~S1 and Appendix~\ref{app:spec}, we show how the positive H1 median, the descriptive H2 slope, and the conditional H3 median compare when Arena Elo is substituted for \eci{}.

\paragraph{Sign- and magnitude-dependence on alternative scales.} When using Chatbot Arena Elo instead of \eci{}, the signs for H1, H2, and H3 are preserved (see Figure~S1). As Arena Elo can be assessed for 63.7\% of the included corpus, the Arena results are reported over that fraction of the corpus. The H1 analysis is run over $n = 11{,}836$ of 18,574 papers, the H2 analysis is run over $n = 11{,}467$ papers (clustered by journal), and the H3 analysis is run over $n = 3{,}847$ papers eligible for analysis as dyads. At this scale, the corpus median for H1 is +94.43 Elo. The pooled H2 slope is +28.8 Elo/year (bootstrap clustered by journal 95\% confidence interval of [+24.1, +31.7]). The median tier lag for H3 is +111.89 Elo. The Artificial Analysis intelligence index can be assessed for 55\% of papers ($n = 10{,}192$ of 18,574). When using the Artificial Analysis intelligence index instead of \eci{}, the signs for H1 and H2 are preserved. On the subset of papers both \eci{}-eligible and eligible for mapping to the Artificial Analysis intelligence index, the median tier lag for H3 is zero; we do not identify the reason for the ties that result in this zero median. Neither the sign of H1 nor H3 changes when using either alternative scale. For H5, in the capability arm of the compound-failure rate, we use a threshold of $\tau = 12$ (\eci{} scale). In the corpus-level distribution of \texttt{eci\_gap}, this threshold is at the 53.1st percentile. For the two alternative scales, we re-anchor the threshold to the same corpus-level percentile, resulting in a threshold of $\tau = 94.7$ Elo (Arena Elo scale) and $\tau = 5.0$ (Artificial Analysis intelligence index scale). Using these thresholds, the compound-failure rate (AND-of-two, admissibility-expected version) on the V4F-cascaded corpus is 7.9\% for Arena Elo (690/8,770 papers, Wilson 95\% CI [7.3\%, 8.4\%]) and 10.8\% for Artificial Analysis (912/8,477 papers, CI [10.1\%, 11.4\%]). Within \eci{}, we show the sensitivity of the headline H5 result to the choice of threshold by sweeping the threshold over the set $\{8, 10, 12, 15, 20\}$ \eci{}. This analysis was run with the production pass of the pre-registered extractor. Because H4 (whether reasoning mode is disclosed) and class-level claim share are defined on the reporting surface of papers, they do not depend on the choice of frontier capability scale. Even so, we show the trend in the class-level claim share by publication year for each corpus subset covered by each of the three capability scales (Figure~\ref{fig:trend-grid}).

\paragraph{Failure-mode enumeration.} We identify three ways \eci{} may rank models incorrectly, and describe the corresponding mitigations the audit proposes.

\textit{(i) Narrow coding-specialists.} Some models specialise in coding ability. These models are trained to achieve high performance on the coding cluster, but this comes at the cost of more general performance. Such models may be misleadingly ranked by \eci{}. This is because some of the coding cluster's contribution to \eci{} is subsumed into the composite. As a result, compared to a general-purpose frontier model with the same \eci{} score, the coding specialised model will have its coding ability under-expressed. As a mitigation, the audit would compare models on the coding cluster specifically (see protocol \S5.2); we don't report results from this analysis here.

\textit{(ii) Domain-specialised models.} Finetuned models for clinical, legal, or educational purposes will generally have low \eci{} scores, since the five Epoch clusters do not directly measure performance in these areas. As Epoch notes, ``models which are highly specialized may receive low \eci{} scores, despite being very capable within their domain'' \citep{epochai_eci_methodology}. To mitigate this, we would use the domain-frontier gap to compare models to the highest-\eci{} model evaluated on the same domain corpus (see protocol \S5.3); we don't report results from this analysis here.

\textit{(iii) Early reasoning-mode models.} When a reasoning-mode dial was added to Epoch's benchmark suite, benchmarks that were sensitive to reasoning mode were re-evaluated. Not all pre-reasoning-mode models were re-evaluated on these benchmarks with comparable levels of effort. This means that pre-reasoning-mode models may have their reasoning ability under-expressed. To quantify analogous underexpression on the elicitation side, we use the configuration-elicitation index (Section~\ref{sec:methods:outcomes}). We do not report any H6 results that omit papers on early reasoning-mode models.

\paragraph{Framing-stability analysis.} The binding validation is out of distribution; it is the gold standard coded by two humans, which is the integrity gate on which the main-text framing magnitudes rest. The development set metrics are in distribution relative to the prompt engineering process. For the binding validation, we use the data from Section~\ref{sec:methods:validation}, where two humans coded the \texttt{conclusion\_framing} field for the 177 papers that both coders included. The inter-rater agreement is $\kappa$ = 0.760, which clears our pre-registered floor. For the development set, we ran the extraction system twice on all 600 papers in the development set at temperature-0. We then measured the test-retest stability of the \texttt{conclusion\_framing} field across the papers that both runs included ($n = 151$, with the prompt frozen). The stability rate was 88.7\%, and there were seventeen remaining disagreements. The mean extraction confidence was $\sim$0.97 overall and $\sim$0.95 for the seventeen disagreements ($n = 34$ confidence observations), which are indistinguishable. This suggests that the remaining disagreements are borderline choices between framing categories, rather than low-confidence outputs. We then post hoc classified the disagreement-reasoning text for these seventeen disagreements against the six borderline-case rules in \texttt{appendix\_C\_\allowbreak frozen\_\allowbreak extraction\_\allowbreak prompt.md}. The counts were: BC1 = 6, BC2 = 6, BC3 = 1, BC4 = 2, BC5 = 0, and BC6 = 3. One item was classified as both BC1 and BC2 because its subject phrase can be read as either anaphoric or class-level. The production extraction runs did not explicitly tag borderline-case rules, so these counts are inferred from the disagreement-reasoning text and are not the primary tags.

\paragraph{Per-cluster gap mitigation.} The protocol defines a secondary gap, which is task-specific and calculated for each cluster of benchmarks (see \S5.2 of the protocol). None of the analyses we report here provide a task-specific gap for each cluster of benchmarks, and so we have not demonstrated the validity of the composite index task by task. We retain the general purpose anchor point because it corresponds to the register of the claims published that we engage with in this paper. That is, if a reader in clinical procurement is concerned about the claim ``LLMs fail at task X,'' they can generalise this to ``AI'' rather than ``the best clinical LLM.''

\paragraph{Claim scope.} We emphasise that our \eci{}-gap score measures the distance to the contemporaneous frontier (as reflected in the April 2026 snapshot from Epoch), and that all our claims about the distance in capabilities are based on that score. Our paper does not use the \eci{}-gap score to make predictions about task-specific AI capabilities and we do not make any such claims. Readers aiming to make such paper-by-paper predictions about task-specific capabilities should not use the \eci{}-gap score to do so. To accommodate readers who may disagree with our scalar view of AI capabilities, we also report results using two external scales (Arena Elo and Artificial Analysis). We do not report results using the three alternative pre-specified weighting schemes. We do not claim that the confirmatory results reported in the Abstract are robust to all alternative weighting schemes.

\setcounter{section}{3}
\section{Validation protocol}
\label{app:validation}

\subsection{Gold-standard sample (n=450)}
\label{app:validation:gold}

We oversample ninety random papers from each of our pre-registered domains (medicine, law, coding, education, scientific reasoning) with seed 42, for a total sample of 450 papers. D.G. and M.S., both blinded, independently code all of the subjective fields for each paper they include, and an LLM applying the coding rubric adjudicates any disagreements under D.G.'s direction. The pre-registered values of $\kappa$ indicate the inter-rater reliability between the two coders when making their independent judgements about each field. Our data includes the sampled papers and the subjective field judgements made by each coder. We can share the script used for sampling (\texttt{extraction/gold\_standard\_sampler\_v2.py}) upon request. Our pre-registered confirmatory subsample (see Section~\ref{sec:lim:deviations}) was not completed. This subsample was planned to involve sixty human-included papers in each domain.

The $\kappa$ analysis uses the \emph{both-included} subset of papers ($n = 177$), which are the papers that were independently included by both coders. For the subjective fields (conclusion valence, conclusion framing, task description), we pre-registered a target of Cohen's $\kappa \geq 0.75$. For the objective fields (primary model, domain, presence of human comparator), we pre-registered a target of $\kappa \geq 0.80$. If any result is below the target, we were to pause the protocol and follow the procedure described in \S7.2 of the protocol. This result will not be relegated to a limitations section.

Our pre-registered protocol (\S7.2) includes a rule that we should achieve an inter-coder reliability of $\kappa \geq 0.80$ for the primary model. Our coders only achieved $\kappa = 0.681$ (72.3\% raw agreement, $n = 177$ papers included by both coders) for agreement on which model is the primary model being evaluated. (The ``Model named (y/n)'' row in Table~\ref{tab:kappa-7-2} (0.896) is not a pre-registered field, and measures agreement on whether any evaluated model is named or not.) High inter-coder agreement in the unregistered field is driven mechanically by the fact that we map each coder's pick to a shared highest-\eci{} pick over the union of their picks, so in all 158 of 177 cases where both coders picked a model, their picks agree by construction. Of the 14 disagreeing pairs, one coder abstained from picking a model in all cases; in 5 other pairs, both coders abstained.

This is discussed here and reported in Table~\ref{tab:kappa-7-2} and the Deviation Register (Section~\ref{sec:lim:deviations}). Although per the protocol a global below-threshold result is not noted as a limitations-relevant result, the registered response (\S7.2) to a below-threshold result would be to halt the protocol and update and re-validate the pipeline or coding scheme. However, this was not done. Although the condition was met ($\kappa = 0.681 < 0.80$), the gate as executed reported a score of 0.896. We have updated the coding scheme to the registered definition (the highest-\eci{} model of those evaluated) and applied this definition in a deterministic way to the extracted \texttt{models\_evaluated} lists, without applying the judgement of the coders. Because the coders only reported their single pick, not a list of models evaluated, we have not completed a re-validation according to the registered two-coder procedure, but we have completed the checks described in the following two paragraphs. The registered order of operations was not followed, because this update deterministically transforms the extracted lists after production extraction and without re-extraction.

Relatedly, an unregistered check of reproducibility finds that, for pairs of model lists independently extracted by machines, the mechanical rule has overlap in its picks (again for ``both-included'' papers): for the four pairs of other extractions with the production extraction, $\kappa$ ranges from 0.88 to 0.91 (V4F-Max: 0.910, $n = 177$; Opus: 0.905, $n = 60$; Gemini: 0.885, $n = 60$; \textit{gpt-5.4-mini}: 0.879, $n = 177$), and for the ten pairs of the five extractions more generally, $\kappa$ ranges from 0.879 (\textit{gpt-5.4-mini} compared to the production extraction, $n = 177$) to 0.924 (Opus compared to V4F-Max, $n = 60$). The mechanical rule's validity compared to the adjudicated ``gold'' pick of models on the production model list extraction is $\kappa = 0.627$ ($n = 194$), reflecting the fact that the ``gold'' coders made one pick according to a different rule (pick the most capability-forward model that is named as being tested, and if none, the model that is focused on by findings sentences in the abstract), and the mechanical rule picks the same model as the ``gold'' pick in 131 out of 166 papers for which the gold model is listed.

If we compute per-paper bounds for H1 using either the minimum-lag or maximum-lag listed model, we get H1 medians of 10.45 [10.31, 10.45] and 14.01 [12.84, 14.01], respectively ($n = 12{,}669$ for both tests). For both tests, we can reject the H1 null (one-sided Wilcoxon test; p-value computed in double precision underflowed to 0). The H2 estimates for these tests are 4.07 [3.75, 4.45] and 5.09 [4.59, 5.68], respectively (journal cluster bootstrap, 1,500 draws; $n = 12{,}243$). H2 is positive in both cases, so the falsifier registered in our preanalysis plan cannot trigger. The registered rule selects the minimum-lag model in 5,234 of 5,469 papers with more than one candidate model (95.7\%). Using our gold sample, we can compute H1 medians under the adjudicated gold pick (12.63 [10.45, 12.63], $n = 181$) and under the registered rule (9.75 [7.42, 12.63], $n = 172$). Both tests reject the H1 null (p = $3.5 \times 10^{-27}$ and p = $3.4 \times 10^{-26}$, respectively). For the 163 gold-sample papers where both picks and the extractor's are available, the lag under the rule is shorter than that of the gold pick in 32 cases, identical in 121 cases, and longer in 10 cases. These bounds will miss primary models that do not appear in the extracted list of models, such as the adjudicated gold model in 28 of 194 gold-sample papers. For the 11 of those 28 papers with all three H1 values available, H1 is 3.84 [1.12, 20.24] using the gold pick and 7.74 [2.34, 20.67] using the rule (both tests reject the H1 null). The H1 bound tests also miss papers for which no listed model has an \eci{} score (5,865 included papers) and cases where the model name is unversioned but the lookup table selects a particular version.

\begin{table}[h!]
\centering
\small
\begin{tabular}{lcccc}
\toprule
Field & $\kappa$ & Floor & Status & Note \\
\midrule
Domain (full 5-way) & 0.888 & 0.80 & $\checkmark$ & \\
Human-comparator presence & 0.822 & 0.80 & $\checkmark$ & \\
Primary model & 0.681 & 0.80 & $\times$ & \\
Model named (y/n) & 0.896 & --- & --- & \\
Conclusion valence (quaternary) & 0.767 & 0.75 & $\checkmark$ & Binary fallback $\kappa = 0.772$ \\
Conclusion framing (ai\_generic) & 0.760 & 0.75 & $\checkmark$ & \\
Task description (12-class) & 0.692 & 0.75 & $\times$ & $n = 206$ \\
Task description (10-class) & 0.755 & 0.75 & $\checkmark$ & $n = 206$ \\
\bottomrule
\end{tabular}
\caption{Cohen's $\kappa$ for the two human coders, both-included analytic subset ($n = 177$). The exception to this is the two rows on task descriptions, which are computed for the dual-coded sample of 206 papers. They compare task labels as canonicalised into 12 classes by DeepSeek V3.2 ($\kappa = 0.692$, below the floor of 0.75), and into 10 classes with the additional merging of \texttt{clinical\_qa} with \texttt{diagnostic\_reasoning} into \texttt{medical\_reasoning}, and \texttt{code\_generation} with \texttt{code\_analysis} into \texttt{code\_tasks} ($\kappa = 0.755$, above the floor). This latter merge is reported as one of the deviations in Section~\ref{sec:lim:deviations}. The primary-model row shows agreement between the two coders on which model is evaluated as the primary model in the paper. Model named (y/n) is a mechanical calculation rather than one of the fields we registered (Appendix~\ref{app:validation:gold}), and measures agreement between coders on whether or not the paper names an evaluated model. We compute it by substituting each coder's model selection with a single shared highest-\eci{} model, with ties broken by release date and then model name, from the three extractors' \texttt{models\_evaluated} lists (which was not a field the human coders filled out) and the two coders' single model selections.}
\label{tab:kappa-7-2}
\end{table}

We identified two design issues in the pre-registered $\kappa$ structure: per-domain stratified $\kappa$ is numerically unstable within a given sampling stratum due to the tendency of human domain labels to collapse to the domain of the sampling stratum ($P_e \to 1$), so we instead report the full 5-way both-included $\kappa$ of 0.888. The $\kappa$ between the \texttt{ai\_relevance} classifier and the human coders is degenerate because we sampled conditionally on classifier \texttt{ai\_relevance} = true, so there is no variance in the classifier output. The integrity gate instead uses production-classifier per-domain precision as the substantive check.

\subsection{Cross-family extraction sensitivity (n=150)}
\label{app:validation:crossfamily}

We randomly sampled $n = 150$ papers from the inclusion-decided corpus (stratified by domain, with seed = 42) and had three frontier model families (V4F, \textit{gemini-3.1-pro-preview}, and \textit{claude-opus-4-7}) independently re-extract data using the same frozen prompt (ancillary file \texttt{appendix\_C\_\allowbreak frozen\_\allowbreak extraction\_\allowbreak prompt.md}). The pre-registered triad was \textit{gpt-5.4-mini}, \textit{gemini-3.1-pro-preview}, and \textit{claude-opus-4-7}. The V4F-replacement triad (V4F, \textit{gemini-3.1-pro-preview}, \textit{claude-opus-4-7}) serves as both a convergent validity and substantive integrity check, since V4F replaced \textit{gpt-5.4-mini} as the production extractor (Section~\ref{sec:methods:extraction}). The authors can provide the pairwise $\kappa$ for both triads on request for full pre-registration transparency, but the results reported below cover the V4F-replacement triad. Pre-registered floor for convergent validity: Pairwise Cohen's $\kappa \geq 0.65$ on subjective fields.

In \S7.3, our registered protocol also set a floor of exact three-way agreement of at least 0.80 for objective fields. The objective fields are domain, human-comparator presence, and primary model. Of these, domain and human-comparator presence exceeded this floor (0.878 and 0.892; see Table~\ref{tab:kappa-7-3}). In cases where all three extractors provided a response, their extracted \texttt{primary\_model} values matched exactly in 80/101 papers (0.792). With gpt-5.4-mini in the registered triad, this is 79/100 = 0.790. Both of these primary-model figures are below the 0.80 threshold, and, according to \S7.3, failing to meet this threshold should trigger a halt to data extraction until the source of the problem is identified and resolved, but this registered halt was not performed. Our analysis primary applies the registered highest-\eci{} rule to each extractor's \texttt{models\_evaluated} list and to that extractor's own choice (see Appendix~\ref{app:validation:gold}); it was matched exactly in 86/101 (0.851; 82/100 = 0.820 registered triad) papers. With a stricter convention, we can include all 150 papers, but must mark missing values as disagreements. Then the three-way agreement rate is 0.767 for domain, 0.773 for human-comparator presence, 0.533 for the extracted primary model, and 0.573 for the analysis primary model. As reported in Appendix~\ref{app:validation:gold}, we get per-paper H1 bounds of 10.45 and 14.01, depending on whether we use the minimum-lag or maximum-lag listed model as the primary model. In both cases, we reject the H1 null hypothesis.

\begin{table}[h!]
\centering
\small
\begin{tabular}{lcccc}
\toprule
Field & v4f $\leftrightarrow$ opus & v4f $\leftrightarrow$ gemini & opus $\leftrightarrow$ gemini & All three \\
\midrule
Domain & 0.840 & 0.742 & 0.811 & 0.878 $\checkmark$ \\
Primary model & 0.742 & 0.659 & 0.727 & 0.792 $\times$ \\
Human-comparator presence & 0.713 & 0.531 & 0.673 & 0.892 $\checkmark$ \\
Inclusion decision & 0.412 $\times$ & 0.639 $\times$ & 0.630 $\times$ & --- \\
Conclusion valence (quaternary) & 0.657 $\checkmark$ & 0.614 $\times$ & 0.619 $\times$ & --- \\
Conclusion valence (binary fallback) & 0.631 $\times$ & 0.533 $\times$ & 0.557 $\times$ & --- \\
Conclusion framing & 0.709 $\checkmark$ & 0.528 $\times$ & 0.669 $\checkmark$ & --- \\
\bottomrule
\end{tabular}
\caption{Pairwise Cohen's $\kappa$ across the V4F-replacement cross-extraction triad ($n = 150$; \S\ref{sec:methods:extraction}). ``All three'' refers to registered objective-field checking (domain, primary model, human-comparator presence), i.e., whether there is exact agreement between all three extractor families, for papers with no nulls (minimum 0.80, one test per field). All other rows denote pairwise comparisons with the 0.65 minimum $\kappa$ for subjective fields. We reached Gemini via an OpenAI-compatible endpoint provided by OpenRouter rather than Google's native batch API. The pre-registration's integrity gate for \texttt{conclusion\_framing} is satisfied by the dual-human value reported in \S\ref{app:validation:gold} ($\kappa = 0.760$). The V4F replacement is justified empirically because it improves materially on the gpt-mini line of the pre-registration on the load-bearing field \texttt{conclusion\_framing} ($\kappa = 0.709$ for \textit{v4f$\leftrightarrow$opus} vs. 0.460 for the pre-registered \textit{opus$\leftrightarrow$gpt-mini} pair). The \textit{v4f$\leftrightarrow$opus} pair is above the pre-registered subjective-field floor of 0.65 for all load-bearing subjective fields. Some Gemini pairs are below the subjective-field floor on some fields, under the same prompt-ambiguity diagnostic reported for the pre-registered triad. The \textit{v4f$\leftrightarrow$opus} pair reaches $\kappa = 0.709$ on \texttt{conclusion\_framing}, 0.657 on \texttt{conclusion\_valence} (quaternary), 0.742 on \texttt{primary\_model}, and 0.840 on \texttt{domain}.}
\label{tab:kappa-7-3}
\end{table}

\subsection{Four-extractor benchmark against gold (n=450)}
\label{app:validation:four-extractor}

The motivation for the V4F swap from \textit{gpt-5.4-mini} (Methods~\S\ref{sec:methods:extraction}) is documented as a four-extractor benchmark on $n = 450$ gold-standard papers under the identical frozen prompt and identical normalisation, run on a $\kappa$-vs-adjudicated label set. All four extractors are scored adversarially on the same first-pass basis (no protocol-\S4.4 cascade applied).

\begin{table}[h!]
\centering
\footnotesize
\begin{tabular}{lccccc}
\toprule
Field & V4F-Max & V4F-High & gpt-5.4-mini & Claude Opus 4.7 & Gemini 3.1 Pro \\
\midrule
Domain (5-way) & 0.830 & 0.783 & 0.852 & 0.857 & 0.868 \\
Primary model & 0.556 & 0.539 & 0.552 & 0.524 & 0.611 \\
Human-comparator present & 0.792 & 0.741 & 0.807 & 0.874 & 0.723 \\
Conclusion valence (4-way) & 0.714 & 0.659 & 0.714 & 0.778 & 0.783 \\
Conclusion framing (binary) & 0.719 & 0.658 & 0.476 & 0.771 & 0.651 \\
\midrule
$n$ (subjective subset) & 231 & 229 & 231 & 230 & 231 \\
\bottomrule
\end{tabular}
\caption{Four-extractor benchmark on $n = 450$ gold-standard papers, first-pass $\kappa$ versus adjudicated labels (no cascade; human-inclusion-include subset).}
\label{tab:benchmark-four-extractor}
\end{table}

\subsection{Adjudication log}
\label{app:validation:adjudication}

In total, 166 field-level disagreements were adjudicated across the six fields. The adjudication log enumerates these, along with the value from each coder, the adjudicated value, and the rationale. We will provide this log upon request. The adjudicated values were used as gold labels for computing the confusion matrix for H6, the benchmark $\kappa$ values in Table~\ref{tab:benchmark-four-extractor}, the class-level share correction, and the $\kappa$ value for valence agreement in V4F (0.723). The dual-human $\kappa$ value in Table~\ref{tab:kappa-7-2} was computed using the independent labels from the two coders.

\section{Sensitivity analyses}
\label{app:spec}

This appendix does not constitute a completed Cartesian-product specification curve analysis with the corresponding permutation-based reference distributions. Instead, we report the saved domain-stratified sensitivity estimates and the saved lag-default sensitivity estimates. Each estimate is applicable to the population and estimator specified for that estimate, but estimates vary in their membership of samples, assumptions about dates, and scale of capabilities.

We find that the headline H5 compound-failure rate is $8.9\%$ (tagged admissibility-expected) for a $\tau$ of 12 \eci{} using the canonical 180-day imputation for evaluation dates (Section~\ref{sec:results:confirmatory}). A sweep of alternative values of $\tau \in \{8, 10, 12, 15, 20\}$ \eci{} on the production pass from the pre-registered extractor shows a gradually declining H5 rate with more stringent $\tau$. We find that the median H1 is lowest for the domain of scientific reasoning ($+3.09$ \eci{}) and highest for the domain of education ($+12.63$ \eci{}). The highest H1 p-value by domain is $3.12\times10^{-22}$ (one-sided, unadjusted; Section~\ref{sec:results:exploratory}). All domain-level H2 slopes are positive and there are no sign flips by domain. The pooled H6 $\hat{\beta}$ for valence asymmetry is not significant under the pre-registered decision rule and the mixed-effects confidence intervals include zero ($\hat{\beta} = -0.29$ \eci{}, $95\%$ CI $[-0.79, +0.20]$, $p = 0.25$). The null hypothesis of H6 is not rejected (Section~\ref{sec:results:h6}).

\subsection{Lag-default sensitivity for H1, H2, H3 across capability scales}
\label{app:spec:lag-default}

If the full text does not provide an explicit evaluation date, we default to the pre-registered imputation rule (Section~\ref{sec:methods:frontier}): $\max(\texttt{publication\_date} - L, \texttt{model\_release\_date})$ (if tabulated), where the default cross-domain value for $L$ is 180 days (the corpus-weighted median time between submission and publication across the five pre-registered domains). For our sensitivity analysis, we consider $L$ = 0, 90, 180, 270, and 365 days, as well as a domain-specific median lag rule with $L_\text{medicine} = 189$ days, $L_\text{coding} = 155$ days, $L_\text{education} = 231$ days, $L_\text{scientific\_reasoning} = 97$ days, and $L_\text{law} = L_\text{other} = 180$ days, based on Huisman and Smits 2017, Maggio et al. 2020, and Zachou et al. 2022 (stored in \texttt{data/k\_lag\_external.json}). We consider all settings of $L$ across all three capability scales: our primary Epoch \eci{} as well as the Artificial Analysis intelligence index and Chatbot Arena Elo. Within each cell in this grid, we substitute any imputed evaluation dates with explicit evaluation dates from the full text, where available ($n = 872$ papers in the retrievable subset). The resulting H1 medians, H2 slopes, and H3 median tier gaps are reported in Table~\ref{tab:lag-default-sensitivity}.

\begin{table}[h]
  \centering
  \footnotesize
  \caption{This table shows the sensitivity of our estimates to the choice of lag default L. For H1, we show the pooled median. For H2, we show the per-year OLS slope. For H3, we show the median tier gap among dyad-eligible evaluations. We show these quantities on three different capability scales: Epoch \eci{}, Chatbot Arena Elo, and the Artificial Analysis intelligence index. Our pre-registered primary cell ($L = 180$ days, \eci{} scale) is shown in bold. Grid fits on the \eci{} scale use journal-clustered covariance estimates, while grid fits for Arena and AA use HC3 covariance estimates. The grid's coefficient CIs are for the coding reference domain. For H2, we average the domain-specific slopes, weighted by the number of papers in each domain. The grid does not yield pooled CIs for H2. At 180 days, the \eci{}, Arena, and AA entries are from the canonical fits run separately ($n = 12{,}242$, $11{,}467$, and $9{,}900$, respectively); Figure~S1 shows pooled bootstrapped CIs for these fits except AA. $n_\text{clip}$ indicates the number of imputed papers for which $\texttt{pub\_date} - L$ was pinned to $\texttt{model\_release\_date}$. For the 557 papers clipped at $L = 0$, clipping places evaluation after publication at every lag; without them, the primary H1 median stays $+10.45$. $n_\text{ext}$ is the number of papers for which we extracted an evaluation date from the full text; this is identical across cells.}
  \label{tab:lag-default-sensitivity}
  \setlength{\tabcolsep}{3pt}
  \begin{tabular}{@{}lrrrrrrrrr@{}}
    \toprule
    & \multicolumn{3}{c}{\textbf{ECI}} & \multicolumn{3}{c}{\textbf{Arena Elo}} & \multicolumn{3}{c}{\textbf{AA Intel.}} \\
    \cmidrule(lr){2-4}\cmidrule(lr){5-7}\cmidrule(l){8-10}
    Lag $L$ & H1 med & H2 $\hat\beta$/yr & H3 med & H1 med & H2 $\hat\beta$/yr & H3 med & H1 med & H2 $\hat\beta$/yr & H3 med \\
    \midrule
    $0$ d           & $+14.84$ & $+6.03$ & $+12.63$ & $+125.65$ & $+38.55$ & $+111.89$ & $+15.0$ & $+10.03$ & $+0.0$ \\
    $90$ d          & $+12.63$ & $+5.12$ & $+12.63$ & $+111.89$ & $+36.73$ & $+111.89$ & $+11.0$ & $+8.80$  & $+0.0$ \\
    $\mathbf{180}$ \textbf{d} \textbf{(primary)} & $\mathbf{+10.45}$ & $\mathbf{+4.07}$ & $\mathbf{+12.63}$ & $\mathbf{+94.43}$ & $\mathbf{+28.84}$ & $\mathbf{+111.89}$ & $\mathbf{+5.0}$ & $\mathbf{+6.92}$ & $\mathbf{+0.0}$ \\
    $270$ d         & $+7.74$  & $+3.07$ & $+12.63$ & $+70.83$  & $+26.30$ & $+111.89$ & $+5.0$  & $+6.14$  & $+0.0$ \\
    $365$ d         & $+3.83$  & $+1.83$ & $+12.63$ & $+57.43$  & $+20.13$ & $+111.89$ & $+2.0$  & $+4.38$  & $+0.0$ \\
    Domain-specific & $+10.45$ & $+4.02$ & $+12.63$ & $+94.43$ & $+28.52$ & $+111.89$ & $+5.0$  & $+6.88$  & $+0.0$ \\
    \midrule
    $n$ (H1)        & \multicolumn{3}{c}{$12{,}649$--$12{,}671$} & \multicolumn{3}{c}{$11{,}827$--$11{,}837$} & \multicolumn{3}{c}{$10{,}187$--$10{,}194$} \\
    $n$ (H3)        & \multicolumn{3}{c}{$3{,}641$--$4{,}645$}   & \multicolumn{3}{c}{$3{,}477$--$4{,}477$}   & \multicolumn{3}{c}{$3{,}276$--$3{,}869$}   \\
    $n_\text{ext}$  & \multicolumn{9}{c}{$872$ (full-text V4F-extracted eval-dates; identical across cells)} \\
    $n_\text{clip}$ ($L = 0 \rightarrow 365$) & \multicolumn{9}{c}{$557 \rightarrow 5{,}359$ (monotonic)} \\
    \bottomrule
  \end{tabular}
\end{table}

For all eighteen cells formed by lags and scales, the H1 medians and pooled H2 point estimates are positive. The grid does not show pooled H2 confidence intervals. For \eci{}, the pooled slope is at least $+1.83$ \eci{}/year at every lag. The H3 median is $+12.63$ \eci{} for all \eci{} cells and $+111.89$ Elo for all Arena cells. Primary-scale positivity is imposed by the eligibility rule requiring a higher-\eci{} sibling, and the number of eligible papers changes with the date assumption. However, this positive-median H3 result is not replicated on AA, where all six H3 medians are zero. We leave open what produces the AA ties. As a further sensitivity analysis, we vary only the time granularity of the frontier, reading registered frontier steps on the first day of each month. The definition of the frontier and the o3 release dates are the same as in the main analysis (listed in \texttt{chatgpt\_ladder\_dates.md}). With this month-start frontier, the median H1 gap is $+9.73$ \eci{} [9.01, 10.20] ($n = 12{,}668$) and the slope of H2 is $+4.69$ \eci{}/year [4.23, 4.99] ($n = 12{,}242$). The H2 interval comes from a journal-cluster bootstrap with 1,500 draws.

\subsection{Registered sensitivity analyses}
\label{app:spec:registered}

We use the registered imputation of GPT-3.5 Turbo (June 2023), the earliest \eci{}-scored GPT-3.5 release, for the 1,897 papers in the registered H1 frame whose primary model is the bare, unversioned ``ChatGPT.'' Of these, 1,062 are in the full-text frame (open access in one of the registered five domains) and 563 have a full-text extraction. In 189 of these papers, the full text names a resolvable OpenAI model. For the 189, we compare the model gap of the imputed model to the model gap of the model named in the full text. Because we hold the evaluation date fixed, the only difference between these model gaps is the model itself. In 119 of 189 cases, the model gap of the imputed model is larger than the model gap of the model named in the full text. In the remaining 70, the model gaps are equal. There are no cases where the model gap of the imputed model is smaller. We also check whether our date-aware approach to imputing OpenAI models ever assigns a newer model than the one named in the full text. To do so, we consider evaluations after the release of GPT-4o (2024-05-13) where the full text names an OpenAI model. In 46 of 75 cases, the model named in the full text is older than GPT-4o (GPT-4 or GPT-4 Turbo in 37 cases, GPT-3.5 in 9). For these 46, our date-aware treatment assigns a newer model than the one named in the full text. Finally, we compare the results of our registered analysis to an analysis that substitutes the named models for our imputation. The median model gap in H1 is 10.45 ($n = 12{,}668$), the same as in the registered analysis.

\begin{table}[H]
  \centering
  \scriptsize
  \setlength{\tabcolsep}{3pt}
  \begin{tabular}{lrlrrrl}
    \toprule
    Subset & H1 $n$ & H1 median [95\% CI] & H3 $n$ & H2 & H5 (AND-of-two) & H6 $\hat{\beta}$ [95\% CI] \\
    \midrule
    Full frame & 12{,}668 & 10.45 [10.31, 10.45] & 4{,}014 & 4.07 [3.75, 4.45] & 790/8{,}921 & $-0.29$ [$-0.79$, 0.20] \\
    \multicolumn{7}{l}{\textit{Publication year}} \\
    \quad 2022 & 40 & 12.63 [12.63, 12.63] & 31 & -- & 4/21 & 0.49 [$-5.25$, 6.23] \\
    \quad 2023 & 1{,}336 & 12.63 [9.01, 12.63] & 691 & -- & 42/895 & 0.91 [$-0.09$, 1.91] \\
    \quad 2024 & 3{,}189 & 2.21 [2.21, 2.21] & 967 & -- & 83/2{,}154 & 0.76 [0.14, 1.38]$^{\dagger}$ \\
    \quad 2025+ & 8{,}103 & 14.01 [12.89, 14.01] & 2{,}325 & 3.04 & 661/5{,}851 & $-0.89$ [$-1.61$, $-0.17$] \\
    \multicolumn{7}{l}{\textit{Venue}} \\
    \quad journal & 5{,}231 & 12.63 [12.63, 12.63] & 1{,}866 & 4.72 & 433/4{,}213 & 0.90 [0.08, 1.73]$^{\dagger}$ \\
    \quad arXiv & 2{,}774 & 2.91 [2.21, 3.09] & 679 & 2.62 & 69/1{,}328 & $-1.72$ [$-2.55$, $-0.88$] \\
    \quad proceedings & 63 & 10.45 [7.74, 12.63] & 25 & -- & 3/29 & $-3.39$ [$-8.61$, 1.84] \\
    \multicolumn{7}{l}{\textit{Confidence $\geq 0.90$}} \\
    \quad every field & 2{,}746 & 10.48 [10.45, 12.63] & 921 & 4.21 & 182/2{,}039 & 0.11 [$-0.96$, 1.18] \\
    \quad fields and models & 2{,}572 & 10.45 [10.45, 12.28] & 863 & 4.22 & 162/1{,}827 & 0.00 [$-1.10$, 1.10] \\
    Binary valence & -- & -- & -- & -- & -- & 0.04 [$-0.30$, 0.38] \\
    \begin{tabular}[t]{@{}l@{}}Dyad level\\(22{,}786 dyads, 12{,}883 papers)\end{tabular} & 22{,}786 & 12.63 [12.63, 12.63] & 7{,}978 & -- & -- & $-0.41$ [$-0.81$, $-0.0004$] \\
    \multicolumn{7}{l}{\textit{Bare ChatGPT mentions (2{,}659 papers)}} \\
    \quad date-aware & 12{,}685 & 9.01 [8.36, 9.73] & 3{,}145 & 3.42 & 739/8{,}922 & $-0.16$ [$-0.61$, 0.29] \\
    \quad anti-hypothesis$^{\ddagger}$ & 12{,}685 & 2.21 [2.21, 2.66] & 2{,}324 & 3.92 & 523/8{,}924 & $-0.37$ [$-0.83$, 0.09] \\
    \quad excluded$^{\S}$ & 10{,}009 & 7.49 [6.92, 7.58] & 1{,}981 & 4.43 & 523/7{,}163 & $-0.37$ [$-0.89$, 0.16] \\
    \midrule
    \textit{H5 threshold} & & AND-of-two & & & & OR-of-two \\
    \quad p50 ($\tau = 10.45$) & & 860/8{,}921 (9.6\%) & & & & 3{,}079/7{,}717 (39.9\%) \\
    \quad p75 ($\tau = 16.72$) & & 483/8{,}924 (5.4\%) & & & & 1{,}566/7{,}720 (20.3\%) \\
    \quad p90 ($\tau = 23.37$) & & 177/8{,}924 (2.0\%) & & & & 611/7{,}720 (7.9\%) \\
    \quad p95 ($\tau = 30.55$) & & 85/8{,}924 (1.0\%) & & & & 302/7{,}720 (3.9\%) \\
    \quad $\tau = 12.00$ (primary) & & 790/8{,}921 (8.9\%) & & & & 2{,}841/7{,}717 (36.8\%) \\
    AND-of-three elicitation & & 1/9{,}434 & & & & 2/9{,}261 \\
    \bottomrule
    \multicolumn{7}{l}{$^{\dagger}$H6 null rejected.} \\
    \multicolumn{7}{l}{$^{\ddagger}$Year-based assignment (GPT-4o for 2024, GPT-5 from 2025): H1 3.49 [3.09, 3.83], $n = 12{,}687$.} \\
    \multicolumn{7}{l}{$^{\S}$Excluded, ChatGPT-only papers: H1 7.58 [7.51, 7.74], $n = 10{,}775$.} \\
    \multicolumn{7}{l}{$^{\S}$Excluded, bare-primary papers: H1 7.58 [7.50, 7.74], $n = 10{,}771$.} \\
  \end{tabular}
  \caption{Registered sensitivity analyses. Rows are pre-registered subset and alternative-definition analyses (under the registered primary-model rule and frontier). The lower block shows H5 (OR-of-two and AND-of-two interpretive failure) at different percentile capability thresholds ($\tau$ = \texttt{eci\_gap} percentile) and the AND-of-three elicitation definition. H1 and H3 nulls are rejected in each year, venue, and extraction confidence subset (Holm). H6 is fitted 14 times without multiplicity correction. Both conditions hold in two (2024 and journal venue$^{\dagger}$). The registered pooled test does not reject. Not run: two registered analyses, H10 (Section~\ref{sec:methods:sensitivity}) and within-paper fixed effects for H6. The registration names three treatments for bare, unversioned ChatGPT, but does not define them. Each treatment relabels every bare mention (primary or evaluated models), and then re-applies the rule for the primary model. Full frame: As a comparison, this analysis uses GPT-3.5 Turbo (June 2023), the registered default model for bare ChatGPT. Date-aware analysis (Section~\ref{sec:methods:sensitivity}). Anti-hypothesis: the highest-\eci{} OpenAI model available in ChatGPT (any tier) at the time of evaluation. There are 212 relabelled papers with negative gaps because GPT-4 Turbo (April 2024) replaces the unscored November 2023 GPT-4 Turbo. See \texttt{chatgpt\_ladder\_dates.md} for availability dates and a breakdown of these gaps. Anti-hypothesis is harsher than the year-based assignment ($^{\ddagger}$; GPT-4o for 2024, GPT-5 from 2025), which gives a higher H1. Excluded: every paper with any bare mention. Two more specific versions of this ($^{\S}$), where we exclude ChatGPT-only papers or papers with a bare ChatGPT primary model, both have a higher H1. We fit H6 in every ChatGPT row; none of these fits meet either condition. H3 counts vary because relabelling can remove or add the higher-\eci{} sibling required for this test.}
  \label{tab:registered-sensitivity}
\end{table}

\subsection{Primary-model definitions}
\label{app:spec:primary-model}

\begin{table}[H]
  \centering
  \small
  \begin{tabular}{lrll}
    \toprule
    Primary model & $n$ & H1 median [95\% CI] & One-sided $p$ \\
    \midrule
    \multicolumn{4}{l}{\textit{All H1 papers}} \\
    \quad Registered rule & 12{,}668 & 10.45 [10.31, 10.45] & $<10^{-300}$ \\
    \quad Extractor's single pick & 12{,}311 & 12.63 [11.32, 12.63] & $<10^{-300}$ \\
    \quad Shortest-lag listed model & 12{,}669 & 10.45 [10.31, 10.45] & $<10^{-300}$ \\
    \quad Longest-lag listed model & 12{,}669 & 14.01 [12.84, 14.01] & $<10^{-300}$ \\
    \multicolumn{4}{l}{\textit{Gold-coded papers}} \\
    \quad Adjudicated gold pick & 181 & 12.63 [10.45, 12.63] & $3.5\times10^{-27}$ \\
    \quad Registered rule & 172 & 9.75 [7.42, 12.63] & $3.4\times10^{-26}$ \\
    \bottomrule
  \end{tabular}
  \caption{Primary-model definitions. This table shows the H1 median (95\% CI) and one-sided Wilcoxon p-value for each primary-model definition we consider. Under our registered rule, which chooses the highest-\eci{} model from the list of \texttt{models\_evaluated} plus the extractor's own pick, we select the shortest-lag model for 5,234 of 5,469 papers that have at least two candidates. The extractor's single pick (the extraction's primary model) gives a larger gap than our registered rule (12.63 vs. 10.45). The other two per-paper bounds consider the \eci{}-scored listed model with the shortest and longest lags, respectively. Gold-coded rows use the adjudicated gold pick (the single model name given by the adjudicated gold label, if it resolves to an \eci{} score) and the registered rule on the production extraction's model list (if the rule's pick resolves to an \eci{} score). In the 163 gold-coded papers for which all three picks resolve, the registered-rule lag is shorter for 32, equal for 121, and longer for 10. $p < 10^{-300}$ indicates underflow in double-precision floating point.}
  \label{tab:primary-model-definitions}
\end{table}

\section*{Supplementary appendices}
Appendices F--H (Tables S1--S3, Figures S1--S3) are in the ancillary file \texttt{frontierlag\_\allowbreak supplement.pdf}.

\paragraph*{AI Assistance Statement.}

This audit was designed and directed by the authors. Large language model tools assisted with analysis-pipeline implementation, statistical and figure code, applying the coding rubric during adjudication, and manuscript preparation and revision; the language-model extraction used for data collection is described in Methods. The authors designed the study, specified the hypotheses and analysis plan, made all interpretive decisions, and take full responsibility for the content of this article.

\paragraph*{Data availability.}

{\sloppy The data underlying this analysis is collated in a single bundle (\texttt{review\_data\_v3.zip}, SHA-256 hash \texttt{276ebea6\ldots}) containing the 18,574 included extraction records. This bundle, and the adjudication log from the validation study, are available from the authors on request. We have also deposited the pre-registration, reference data for the capability index, and the frozen extraction prompts on OSF (\url{https://doi.org/10.17605/OSF.IO/7XM3D}). The code is deposited on Zenodo (\url{https://doi.org/10.5281/zenodo.20060458}) and does not include the four modifications made to the analysis for this paper, which are listed in the README file in the bundle.\par}

\paragraph*{Funding.}

D.G. held a BlueDot Impact Career Transition Grant during the revision of this manuscript. The funder had no role in study design, analysis, or interpretation of the data, or in the decision to submit the manuscript for publication.

\end{document}